\documentclass[11pt,document,nofootinbib,superscriptaddress,onecolumn,preprintnumbers,balancelastpage]{article}

\pdfoutput=1
\usepackage{jheppub}


\allowdisplaybreaks

\usepackage{amsmath, amsfonts, amsthm, amssymb}
\usepackage[normalem]{ulem}

\usepackage[toc,page]{appendix}

\usepackage{bm}
\usepackage{upgreek}
\usepackage{comment}
\usepackage{ifpdf}
\usepackage{xstring}
\usepackage{setspace}
\usepackage[utf8]{inputenc}
\usepackage{hyperref}
\usepackage{graphicx}  
\usepackage{breqn}
\usepackage{mathrsfs}

\DeclareMathAlphabet{\mathbbold}{U}{bbold}{m}{n}

\newcount\Comments  
\usepackage{color}
\definecolor{colorGH}{rgb}{.2,.7,.2}
\newcommand{\kibitz}[2]{\ifnum\Comments=1\textcolor{#1}{#2}\fi}

\newcommand{\kk}{\hat{k}}

\newcommand{\be}{\begin{equation}}
\newcommand{\ee}{\end{equation}}
\newcommand{\bea}{\begin{eqnarray}}
\newcommand{\eea}{\end{eqnarray}}

\newcommand{\SLC}{\textrm{SL(2,}\mathbb{C})}

\newcommand{\eq}[2][]{%
    \ifx&#1&
      \begin{equation*} #2 \end{equation*}
    \else
  \begin{equation}\label{#1}{#2} \end{equation}
\fi
}
\newcommand{\eqs}[2][]{%
    \ifx&#1&
      \begin{eqnarray*} #2 \end{eqnarray*}
    \else
  \begin{subequations}\label{#1}\begin{eqnarray} #2 \end{eqnarray}\end{subequations}
\fi
}

\preprint{}

\makeatletter
\renewcommand*{\p@section}{\S\,}
\renewcommand*{\p@subsection}{\S\,}
\makeatother

\title{The Newman-Penrose Map and the  Classical Double Copy}

\author[a]{Gilly Elor,}
\author[b]{Kara Farnsworth,}
\author[c]{Michael L. Graesser,}
\author[d]{and Gabriel Herczeg}

\affiliation[a]{\footnotesize Department of Physics, University of Washington, Seattle, WA 98195, USA}
\affiliation[b]{\footnotesize CEICO, Institute of Physics, Czech Academy of Sciences, 182 21, Prague 8, Czech Republic}
\affiliation[c]{\footnotesize Theoretical Division, Los Alamos National Laboratory, Los Alamos, NM 87545, USA}
\affiliation[d]{\footnotesize Brown Theoretical Physics Center, Department of Physics, Brown University, Providence, RI 02912, USA}

\emailAdd{gelor@uw.edu}
\emailAdd{kmfarnsworth@gmail.com}
\emailAdd{michaelgraesser@gmail.com}
\emailAdd{gabriel\_herczeg@brown.edu}

\preprint{LA-UR-20-23506}
\abstract{
Gauge-gravity duality is arguably our best hope for understanding quantum gravity. Considerable progress has been made in relating scattering amplitudes in certain gravity theories to those in gauge theories---a correspondence dubbed the \emph{double copy}. Recently, double copies have also been realized in a classical setting, as maps between exact solutions of gauge theories and gravity. We present here a novel map between a certain class of real, exact solutions of Einstein's equations and self-dual solutions of the flat-space vacuum Maxwell equations. This map, which we call the \emph{Newman-Penrose map}, is well-defined even for non-vacuum, non-stationary spacetimes, providing a systematic framework for exploring gravity solutions in the context of the double copy that have not been previously studied in this setting. To illustrate this, we present here the Newman-Penrose map for the Schwarzschild and Kerr black holes, and Kinnersley's photon rocket.}

\begin{document}
\maketitle
\newpage
\section{Introduction}
\label{sec:intro}
The gauge-gravity double copy relates perturbative scattering amplitudes in non-abelian gauge theory to amplitudes of a gravitational theory by replacing color factors with additional kinematic information \cite{Bern:2008qj,Bern:2010yg,Bern:2010ue,Bern:2019prr}. Originally formulated to relate open and closed string amplitudes via the KLT (Kawai-Lewellen-Tye) relations \cite{kawai1986relation}, this subject has seen great progress in recent years: color-kinematic 
duality and the double copy have been proven at tree level \cite{Bern:2010yg}, appear to hold at loop level \cite{Bern:2010ue,BjerrumBohr:2012mg,Nohle:2013bfa,Boels:2013bi,Bern:2013yya,Du:2014uua}, and are widely believed to hold to all orders in perturbation theory \cite{Oxburgh:2012zr,Naculich:2013xa}. The power of the double copy in relating gravity to gauge theories provides new computational tools that can, for instance, be leveraged to simplify and improve calculations of black hole merger in-spirals \cite{Cheung:2018wkq,Bern:2019crd,Bern:2019nnu,Leibovich:2019cxo}.

Beyond computational advantages, the double copy offers a new perspective on the deep mathematical relationship between gauge theories and gravity, which could provide crucial insights toward  developing a consistent theory of quantum gravity. 
In order to make progress in this direction, one may wonder if a double copy prescription exists beyond the amplitude level. While there has been some work hinting at a Lagrangian-level realization \cite{Cheung:2016say, Cheung:2017kzx}, a large body of recent work focuses on developing systematic procedures to map between certain classes of classical solutions of gauge theory and gravity, and studying explicit examples of such dual solutions.
Many of the examples studied under this classical double copy have simple and intuitive counterparts; for instance, the Schwarzschild and Kerr black hole solutions of general relativity can be mapped to solutions of the vacuum Maxwell equations sourced by a point charge and a rotating disk of charge respectively \cite{Monteiro:2014cda} (for an overview see \cite{White:2017mwc}).

One may also reverse this logic and perform a ``single copy" to obtain gauge amplitudes by replacing color information with kinematic information in a bi-adjoint scalar field theory---an interacting scalar theory in which the scalar transforms in the adjoint of two distinct Lie groups \cite{BjerrumBohr:2012mg,Du:2011js}. Despite having no known phenomenological realizations, it is becoming clear that such bi-adjoint scalar theories play an important role in studies of the perturbative double copy and 
color-kinematic duality \cite{Bern:2008qj}, and furthermore correspond to one node in a web of theories interconnected at the perturbative level \cite{Cheung:2017ems}. On the classical side, progress has been made by  identifying exact (non-perturbative) solutions of the bi-adjoint scalar theory and comparing their properties to monopole solutions in non-abelian gauge theory \cite{White:2016jzc,DeSmet:2017rve,Bahjat-Abbas:2018vgo}.

While an impressive breadth of literature exists which involves discovering particular examples that display double and single copy behavior, one is left with several lingering questions on the generality of this prescription. Crucially, 
the classical double copy prescription appears to be manifest generally only in \emph{stationary} solutions\footnote{Recall that a spacetime is said to be \emph{stationary} if it admits a Killing vector that is timelike in a neighborhood of spatial infinity.} of the \emph{vacuum} Einstein equations that can be expressed in Kerr-Schild form \cite{article}.\footnote{\footnotesize{See \cite{Carrillo-Gonzalez:2017iyj} for work exploring contracting more general Killing vectors with the Einstein equations.
}}
It would therefore be desirable to unravel the conditions linking the success of the classical double copy to the Kerr-Schild form, and to search for possible ways to extend the double copy to more general spacetimes. There are also salient examples of simple exact solutions in gauge theory, e.g. electromagnetic dipoles and Yang-Mills instantons, that have no obvious dual solutions on the gravity side using current techniques. 

For the perturbative double copy, progress has been made to include an analysis of bound state systems \cite{Goldberger:2017vcg}; by studying the radiation field sourced by such systems, color dipoles were identified with a gravity quadrupole moment. Headway has also been made in understanding the classical double copy by considering solutions that are self-dual in both the gauge and gravitational theories \cite{Monteiro:2014cda, Berman:2018hwd}---in the self-dual double copy prescription, one formulates the double copy in terms of a differential operator which generates gauge and gravity solutions from a harmonic function of the bi-adjoint scalar theory. This prescription allowed the authors of \cite{Berman:2018hwd} to identify a dipole-like self-dual gauge field with the Eguchi-Hanson gravitational instanton 
\cite{Eguchi:1978xp}.
 The classical double copy prescription was also extended to double Kerr-Schild spacetimes by \cite{Luna:2015paa}, where it was shown that a dyon on the gauge theory side can be double copied to the Taub-NUT solution on the gravity side. Furthermore, the classical double copy can also be applied to generalized Kerr-Schild metrics expanded around a curved background metric, where the dual gauge field solves the vacuum Maxwell equations on the curved background \cite{Bahjat-Abbas:2017htu}. Although the classical double copy has been shown to hold for certain specific non-stationary solutions, such as PP wave spacetimes \cite{Luna:2018dpt} and Kinnersley's accelerating and radiating black hole metric \cite{Kinnersley:1969zz, Luna:2016due}, a systematic approach to studying the classical double copy for non-stationary spacetimes has heretofore been lacking, except when the spacetime is of Petrov type D or N \cite{Luna:2018dpt}. Double copy prescriptions for non-vacuum spacetimes have also been proposed---where for instance, the stress energy tensor on the gravity side is related to the ``square" of a gauge current \cite{Luna:2016due,Bah:2019sda}, however such prescriptions can lead to inconsistencies.\footnote{The source terms introduced in \cite{Monteiro:2014cda} cannot simultaneously satisfy the weak and strong energy conditions \cite{Ridgway:2015fdl}.} For another approach to studying non-vacuum solutions in the double copy see \cite{Kim:2019jwm}.

In this work we introduce the \emph{Newman-Penrose map}---a novel map, closely related to the classical double copy, that associates a self-dual solution of the vacuum Maxwell equations to certain Kerr-Schild spacetimes that need be \emph{neither stationary nor pure vacuum}.  We make only the mild assumptions that the Kerr-Schild vector is tangent to a shear-free, null geodesic congruence (SNGC),  and that this congruence has non-vanishing expansion (see \ref{sec:optic} for a review of the optical properties of such congruences). 

The assumption that the Kerr-Schild vector be geodesic is standard; the Kerr-Schild vector $k^\mu$ is geodesic if and only if the stress tensor obeys the condition $T_{\mu\nu} k^\mu k^\nu = 0$, which includes both vacuum spacetimes and spacetimes with a pure radiation stress tensor $T_{\mu\nu} \propto k_\mu k_\nu$. Moreover, the assumption that $k^\mu$ is geodesic implies that it is also a repeated principal null direction of the Weyl tensor, so that the spacetime is algebraically special \cite{Stephani:2003tm}. 
If, additionally, the vacuum Einstein equations are imposed, then the Goldberg-Sachs theorem ensures that the Kerr-Schild vector must be shear-free  \cite{goldberg2009republication}. Since we do not want to restrict ourselves to vacuum spacetimes, we impose the shear-free condition explicitly. Among vacuum Kerr-Schild spacetimes with a geodesic Kerr-Schild vector, the restriction that the expansion of the congruence generated by the Kerr-Schild vector be non-vanishing excludes only a small subset of vacuum Kerr-Schild spacetimes consisting of certain exact gravitational wave solutions---in particular, it excludes exactly the solutions of Petrov type N, but includes the much larger class of solutions of Petrov type II or D \cite{mcintosh1989kerr, mcintosh1988single, Stephani:2003tm}. Among the Kerr-Schild spacetimes with an expanding SNGC, there are also non-vacuum solutions of physical interest, including Kinnersley's ``photon rocket"---an exact solution of Einstein's equations with a pure radiation stress tensor that describes the gravitational field produced by an arbitrarily accelerating massive particle \cite{Kinnersley:1969zz}.

Under the Newman-Penrose map, the approach taken in the self-dual classical double copy can be extended to a broad class of non-self-dual Kerr-Schild spacetimes. In particular, associated with every Kerr-Schild spacetime admitting an expanding SNGC, there exists a harmonic complex scalar field $\Phi$ such that $A = \hat{k}\Phi$ is a self-dual solution of the vacuum Maxwell equations, where $\hat{k} = dx^\mu\hat{k}_\mu $ is a universal, one-form valued operator, whose origin may be understood from the spinorial realization of the Newman-Penrose map.
This prescription relies on elements of the Newman-Penrose formalism---the natural formalism for constructing algebraically special spacetimes in four dimensions. Note that in this construction we do not require that $\Phi$ satisfies the Pleba\'{n}ski equation for self-dual gravity, and consequently, we necessarily lose the explicit double copy nature of the graviton, i.e. $h_{\mu\nu} \neq \hat{k}_\mu \hat{k}_\nu \Phi$. However, we gain a precise map that can be applied to a more general class of spacetimes which reproduces and extends upon known classical double copy results.

This paper is organized as follows: 
In \ref{sec:clDC}, we review the recent progress in relating classical solutions of gauge theory and gravity via the double copy program, paying particular attention to the \emph{Kerr-Schild} and \emph{self-dual} double copies. In \ref{sec:NPforKS} we review the Newman-Penrose formalism  \cite{Newman:1961qr} for Kerr-Schild spacetimes with an expanding SNGC, largely based on \cite{mcintosh1988single}.
In \ref{subsec:generalProcedure} we introduce the Newman-Penrose map, which we then illustrate by studying the examples of the Schwarzschild and Kerr black holes and Kinnersley’s photon rocket in ~\ref{sec:examples}. In ~\ref{subsec:ManyToOne} we comment on the many-to-one nature of the Newman-Penrose map, emphasizing that information contained in solutions on the gravitational side that does not have an analog on the gauge theory side is ``projected out": for instance, the electric charge in electro-vacuum solutions of the Einstein-Maxwell equations, or the Bondi mass aspect in solutions with a pure radiation stress tensor. We conclude in ~\ref{sec:discussion} with a discussion of our main results and some possible directions for future work. Appendices are included. In \ref{sec:KerrThm} we provide a brief review of the spinorial formalism for general relativity, which we use to summarize the proof of Kerr's Theorem---a crucial result for the construction of algebraically special Kerr-Schild spacetimes. In \ref{app:khat} we provide a spinorial realization of the Newman-Penrose map, giving insight into the origin of the operator $\hat{k}$, which appears somewhat mysterious in the tensorial approach.  In \ref{app:ConfirmMetric} we confirm the Schwarzschild and Kerr metrics can be generated from the $\Phi$ originally presented in \cite{mcintosh1988single}, albeit in a
different coordinate system from the usual double copy starting point.

\newpage
\section{The Classical Double Copy}
\label{sec:clDC}
In this section we briefly review the Kerr-Schild classical double copy, and summarize the result for Schwarzschild spacetime. Additionally, we review the zeroth and single copy relating solutions in gauge theory to bi-adjoint scalar theory. Finally, we include a summary of the self-dual double copy \cite{Monteiro:2014cda}, the formalism of which parallels some of the framework introduced here. 

\subsection{The Kerr-Schild Double Copy}\label{KSDC}
Kerr-Schild solutions are a class of solutions to Einstein's equations that can be written in the form 
\vspace{-0.1 in}
\begin{align}
\label{eq:metricKS}
g_{\mu \nu} = \eta_{\mu \nu} -  \varphi k_\mu k_\nu\,,
\end{align}
where $\eta_{\mu \nu}$ is a flat metric with signature\footnote{Note that we use the mostly minus sign convention for the metric which is typical of the Newman-Penrose formalism, though unusual in the context of the double copy and much of the general relativity literature. This will introduce relative minus signs in many formulae familiar from the classical double copy literature (e.g. the sign in front of $\varphi$ in eqn. \eqref{eq:metricKS}).} $(+,-,-,-)$ (though we do not necessarily use Cartesian coordinates), $\varphi$ is a real function, $k^\mu$ is null with respect to both $g_{\mu \nu}$ and $\eta_{\mu \nu}$ and geodesic $k^\mu \partial_\mu k^\nu = k^\mu \nabla_\mu k^\nu = 0$ with respect to either metric. In general, $k^\mu$ is tangent to a null geodesic congruence if and only if the stress tensor obeys \cite{Stephani:2003tm}
\be
T_{\mu\nu}k^\mu k^\nu = 0\,, 
\label{eq:TKSform}
\ee
which includes vacuum spacetimes with $T_{\mu\nu} = 0$, pure radiation spacetimes with $T_{\mu\nu} = f k_\mu k_\nu$, as well as null and non-null   electro-vacuum solutions\footnote{\footnotesize{We comment on the novel treatment of such solutions e.g. Kerr-Newman black holes,  under the Newman-Penrose map in ~\ref{subsec:ManyToOne}.}}.
With these assumptions, the Ricci tensor with mixed indices truncates at linear order in graviton $h_{\mu\nu} \equiv \varphi k_\mu k_\nu$:
\begin{align}
R^\mu_{\ \nu} = \frac{1}{2}\partial_\rho\left[\partial_\nu h^{\mu\rho} + \partial^\mu h_\nu^{\ \rho} - \partial^\rho h^{\mu}_{\ \nu}\right],
\end{align}
where we denote by $\partial_\mu$ the covariant derivative operator associated with $\eta_{\mu\nu}$ and we have defined $\partial^\mu \equiv \eta^{\mu\nu}\partial_\nu$. For stationary spacetimes admitting a timelike killing vector, 
Einstein's equations then reduce to (where we have chosen the normalization $k_0 = 1$) 
\begin{align}
R^\mu_{\ 0} = \frac{1}{2}\partial_\rho\left[ \partial^\mu (\varphi k^\rho) - \partial^\rho (\varphi k^\mu)\right] = 0\,,
\end{align}
which are exactly the  vacuum equations of motion (Maxwell's equations) ${\partial_\mu F^{\mu \nu} = 0}$, for an abelian gauge field $A_\mu \equiv \varphi k_\mu$ and, when dressed with a color factor $c^a$, can be interpreted as the linearized Yang-Mills equations for a non-ablelian gauge field $A^a_\mu = c^a \varphi k_\mu$. The prototypical example of this correspondence relates the Schwarzschild black hole to a Coulomb field. 

Written in Kerr-Schild form and using spherical coordinates, the Schwarzschild solution is
\begin{align}
\label{eq:Sk}
g_{\mu\nu} &= \eta_{\mu\nu} - \varphi  k_\mu k_\nu \,,\\ 
k_\mu dx^\mu &= dt + dr\,, \nonumber
\end{align}
where $\varphi = 2GM/r$, which corresponds to the line element
\begin{align}
\label{eq:Smetric}
ds^2 = \left( 1 - \frac{2GM}{r} \right) dt^2 - \frac{4GM}{r} dt \, dr - \left(1 +  \frac{2GM}{r} \right) dr^2 - r^2 d \Omega^2\, .
\end{align}
Taking the single copy of this solution and making the replacements, $8\pi G \to 1/\epsilon_0$, $M \to Q$ gives, after a gauge transformation,
\begin{align}
A = A_\mu dx^\mu = \frac{Q}{4\pi\epsilon_0 r}dt\,,
\end{align}
which is a Coulomb field. This prescription has been tested for many different Kerr-Schild spacetimes (see \cite{White:2017mwc} for a review), usually mapping a particular mass distribution on the gravitational side to an intuitively similar charge distribution on the gauge side. The example we have just presented is particularly striking: the gravitational field due to a spherically symmetric mass distribution or black hole maps directly onto the electric field of a spherically symmetric charge distribution or point charge. 

\subsection{The Self-Dual Double Copy}
\label{subsec:SelfDualDC}
There has also been progress in studying the double copy for \emph{self-dual}  solutions in gauge theory and gravity \cite{Monteiro:2014cda, Berman:2018hwd}---i.e. complex solutions with a single physical degree of freedom for the photon or graviton characterized by the conditions\footnote{Where $\star\star \Omega = (-1)^{s + p(n-p)} \Omega$ for any $p$-form $\Omega$ in $n$-dimensional spacetime with signature $s$. Note that defining self-dual vs. anti self-dual solutions in Lorentzian signature is ambiguous up to a sign \cite{Baez:1995sj}, and we will use ``self-dual" to refer to either choice of sign.}
\be
F = \pm\, i\star_0 F\,, \qquad C_{ab} = \pm\, i\star C_{ab}\,,
\ee
\noindent where $\,\star_0\,$ is the Hodge star operator associated with the flat metric $\eta_{\mu\nu}$, $\star$ is the Hodge star operator of the dynamical spacetime metric $g_{\mu\nu}$, $F = dA$ is the field strength two-form associated with the gauge field $A$, and $C_{ab} = C_{ab\mu\nu}\, dx^\mu\wedge dx^\nu$ is the Weyl two-form written in an arbitrary basis $e^a_\mu$.  
In the self-dual  double copy, the null vector $k^\mu$ is promoted to a differential operator: 
\begin{align}
g_{\mu \nu} = \eta_{\mu \nu} -2 \hat{k}_\mu \hat{k}_\nu (\phi) , \qquad g^{\mu \nu} = \eta^{\mu \nu} +2 \hat{k}^\mu \hat{k}^\nu (\phi) ,
\label{eq:MetricSelfDual}
\end{align}
where $\phi$ is a complex scalar field. Symmetry of the graviton requires that the operator commutes with itself $\bigl[ \hat{k}_\mu, \hat{k}_\nu \bigr] = 0$ and the null and geodesic conditions correspond to $\eta_{\mu \nu} \kk^\mu \kk^\nu (\psi) = \hat{k}^2 (\psi) = 0$ and $(\hat{k}\cdot \partial) \psi = 0$ respectively, where $\psi$ is an arbitrary function. The vacuum Einstein equations, $R_{\mu \nu} = 0$, then reduce to a single equation for the scalar $\phi$: 
\be
\partial^2 \phi +  (\kk^\mu \kk^\nu \phi)(\partial_\mu \partial_\nu \phi) =0\,,
\ee
which when written in light-cone coordinates $(u,v,\zeta, \bar{\zeta})$ (defined below in \eqref{eq:coor}) reproduces the Pleba\'{n}ski equation \cite{Plebanski:1975wn,Sabharwal:2019ngs,Monteiro:2014cda, Berman:2018hwd}
\be
\phi_{,vu} - \phi_{,\zeta \bar{\zeta}} = (\phi_{,u\zeta})^2 - \phi_{,\zeta\zeta} \phi_{,vv}
\label{eq:PlebEq}
\ee
for self-dual gravity. Finally, demanding that the graviton can be factorized into the Kerr-Schild form $h_{\mu\nu} \propto k_\mu k_\nu$ leads to an additional constraint equation on $\phi$, which together with the Pleba\'{n}ski equation, implies that $\phi$ is harmonic with respect to the flat-space D'Alembertian $(\partial^2 \phi = 0)$. This condition can be interpreted as a linearized equation of motion for a biadjoint scalar $\Phi^{aa'} = c^a \tilde{c}^{a'} \phi$ with trivial color dependence $c^a$, $\tilde{c}^{a'}$. It is then straightforward to show that $A^a_\mu = c^a\hat{k}_\mu\phi$ and $h_{\mu\nu} = \hat{k}_\mu\hat{k}_\nu \phi$ are self-dual solutions of the Yang-Mills and Einstein equations. In \ref{subsec:generalProcedure}, we will reproduce explicitly the fact that $A = \hat{k}\phi$ with $\hat{k} = dx^\mu \hat{k}_\mu$ is a self-dual Maxwell field, i.e. $F =\pm  i \star_0 F$ whenever $\phi$ is harmonic with respect to the flat-space D'Alembertian $(\partial^2 \phi = 0)$.

The Newman-Penrose map presented in this work will follow a similar construction as the one reviewed here. We will find that $\kk$ yields a self-dual gauge field when acting upon a harmonic function associated with a Kerr-Schild spacetime. However, as stated above, we will not use the self-dual graviton definition $h_{\mu\nu} = \hat{k}_\mu\hat{k}_\nu \phi$ and will instead consider more general spacetimes.

\section{Newman-Penrose Formalism for Kerr-Schild Spacetimes}
\label{sec:NPforKS}

In this section we present a brief overview of Kerr-Schild spacetimes in the Newman-Penrose formalism, following closely the works of Kerr and Wilson \cite{1979GReGr..10..273K}, McIntosh \cite{mcintosh1989kerr} and McIntosh and Hickman \cite{mcintosh1988single} (see also chapter 1 of
\cite{Chandrasekhar:1985kt}, and chapter 32 of \cite{Stephani:2003tm}). 
 We make a few mild assumptions, but the considerations in this section are otherwise quite general. First, we restrict our attention to real, Lorentzian spacetimes of dimension four. Second, we assume that the Kerr-Schild vector $k^\mu$ is tangent to an SNGC. While this condition sounds quite strong, a review of the literature reveals that most Kerr-Schild spacetimes of interest are of this type. When the Kerr-Schild vector $k^\mu$ is tangent to a geodesic congruence, it can be shown (see theorem 32.3 of \cite{Stephani:2003tm}) that the spacetime is algebraically special, with $k^\mu$ being a repeated principal null direction of the Weyl tensor\footnote{Recall that the Weyl tensor of any four-dimensional spacetime yields a set of four null eigenvectors referred to as \emph{principal null directions.} If at least two of these directions coincide, the spacetime is said to be algebraically special, with the corresponding direction(s) referred to as ``repeated principal null direction(s)." }.
 Finally, we will assume that the expansion scalar of the SNGC is non-vanishing. Within the class of vacuum Kerr-Schild solutions, this last assumption excludes exactly the solutions of Petrov type N, which, alas, includes some interesting exact gravitational wave solutions, e.g., PP waves.

\subsection{Metric, Null Coordinates, and Null Tetrad}

We adopt coordinates $(u,v,\zeta,\bar{\zeta})$, where $u,v$ are real light-cone coordinates, and $\zeta,\bar{\zeta}$ are complex conjugate coordinates related to the usual Cartesian coordinates $(t,x,y,z)$ by 
\be
\label{eq:coor}
u = \tfrac{1}{\sqrt{2}}(t - z)\,, \qquad v = \tfrac{1}{\sqrt{2}}(t+ z)\,, \qquad \zeta = \tfrac{1}{\sqrt{2}}(x + iy)\,, \qquad \bar{\zeta} = \tfrac{1}{\sqrt{2}}(x - iy)\,. 
\ee

In these coordinates, one can write a null tetrad  for any Kerr-Schild metric with an expanding SNGC, $l^\mu$, which is proportional to $k^\mu$ of the Kerr-Schild double copy. In this language, a Kerr-Schild metric is one that can be written in the form\footnote{$V$ and $\bm{\omega}^2$ are related to $\varphi$ and $k_\mu$ of section \ref{KSDC} up to overall normalizations.}
\begin{equation}
g = g_0 + V\bm{\omega}^2 \otimes \bm{\omega}^2, \label{VKS}
\end{equation} 
where $g_0$ is a flat metric and $\bm{\omega}^2$ is a null one-form with $l^\mu = g^{\mu\nu}\bm{\omega}^2_\nu$, denoted as such since it will be identified with the second element of a preferred set of null basis one-forms 
\begin{eqnarray}
\bm{\omega}^1 &=& dv + \tfrac{1}{2}V\bm{\omega}^2 ,\nonumber\\
\bm{\omega}^2 &=&  du+ \bar{\Phi}d\zeta + \Phi d\bar{\zeta} + \Phi\bar{\Phi}dv\, , \nonumber\\
\bm{\omega}^3 &=&   \Phi dv + d\zeta\nonumber\,, \\
\bm{\omega}^4 &=& \bar{\Phi}dv + d\bar{\zeta}\,,
\label{eq:tetrad}
\end{eqnarray}
where $V$ is a real function and $\Phi$, $\bar{\Phi}$ are complex conjugates\footnote{When the spacetime is complex, $\Phi$ and $\bar{\Phi}$ may be regarded as independent complex functions. However, in this article we will restrict our attention to real spacetimes.}\textsuperscript{,}\footnote{Our definitions differ from \cite{Stephani:2003tm}, chapter 32 by the exchange $\{\bm{\omega}^1, \bm{\omega}^2\}\leftrightarrow \{\bm{\omega}^3, \bm{\omega}^4\}$  due to our choice of signature.}. The full metric is determined completely once $V$ and $\Phi$ are specified. Although $V$ satisfies its own potentially interesting set of constraints (see \cite{Stephani:2003tm}), from now on we focus on the complex function $\Phi$, which will turn out to be harmonic with respect to the background metric and so an intuitive starting point for mapping gravitational solutions to self-dual gauge fields. For vacuum spacetimes $V$ is completely determined in terms of $\Phi$ and the mass $M$.\footnote{For the Schwarzschild solution, see \eqref{V-Schwarzschild}.}

The form of \eqref{eq:tetrad} can be understood from Kerr's Theorem \cite{Penrose:1967wn, 1976CMaPh..47...75C}, which gives the most general form of an SNGC in Minkowski space (see \ref{sec:KerrThm} for more details in the spinorial representation). The result of the theorem states that the dual of the tangent vector of any such SNGC in Minkowski space is either equal to $dv$, in which case it has vanishing expansion, or it is of the form $\bm{\omega}^2$ in \eqref{eq:tetrad}, where the complex function $\Phi$ is defined implicitly through the vanishing of an arbitrary function $F$ that is analytic in 
three variables \cite{Penrose:1967wn,1976CMaPh..47...75C, Huggett:1986fs}
\begin{equation}
    F(\Phi, u + \Phi\bar{\zeta}, \zeta + \Phi v)=0\, .
    \label{Eqn:KerrTheorem}
\end{equation}

The optical scalars of $l^\mu$---the expansion, shear and twist---are the same whether measured with respect to the flat background metric $g_0$ or the full spacetime metric $g$ (see section \ref{sec:optic} for details). Moreover, it can be shown that $l^\mu$ is geodesic with respect to $g_0$ if and only if it is geodesic with respect to $g$. Taken together, these facts imply that $\bm{\omega}^2$ is dual to an expanding SNGC of the full spacetime metric $g$. 

The full metric can then be written in terms of the basis one-forms \eqref{eq:tetrad} as 
\begin{equation}
g = 2(\bm{\omega}^1\bm{\omega}^2 - \bm{\omega}^3\bm{\omega}^4)\,, \label{full}
\end{equation}
which has the form \eqref{VKS} with the flat metric
\begin{equation}
g_0 = 2(dudv - d\zeta d\bar{\zeta})\,. \label{background}
\end{equation}
We also define the null  tetrad vectors
\bea
l &=& \partial_v - \Phi\partial_\zeta - \bar{\Phi}\partial_{\bar{\zeta}} + \Phi\bar{\Phi}\partial_u\, , \nonumber \\
n &=& \partial_u - \tfrac{1}{2} Vl\, , \nonumber\\
m &=& \partial_{\zeta} - \bar{\Phi}\partial_u \nonumber \,, \\
\bar{m} &=& \partial_{\bar{\zeta}} - \Phi\partial_u\, ,
\label{tetrad vectors}
\eea
where $l = l^\mu\partial_\mu$, etc., and we denote the covariant directional derivative operators associated with these basis vectors by
\bea
D &=& l^\mu\nabla_\mu \,, \qquad \Delta = n^\mu\nabla_\mu\, , \qquad \delta = m^\mu\nabla_\mu\, , \qquad \bar{\delta} = \bar{m}^\mu\nabla_\mu\, . 
\label{directional-derivatives}
\eea

Given the metric $g$ in equation \eqref{full}, one can check that the null tetrad \eqref{tetrad vectors} satisfies the standard normalization conditions 
\be
l^\mu n_\mu = 1 = - m^\mu\bar{m}_\mu\, ,
\label{eq:tentradnorm}
\ee
with all other contractions vanishing\footnote{Note that with these conventions we have 
\be
\bm{\omega}^1_\mu = n_\mu\, , \qquad \bm{\omega}^2_\mu = l_\mu\, , \qquad \bm{\omega}^3_\mu = -\bar{m}_\mu\, , \qquad \bm{\omega}^4_\mu = -m_\mu\, .
\ee}, and that we may write the inverse metric as 
\be
g^{\mu\nu} = l^\mu n^\nu + n^\mu l^\nu - m^\mu \bar{m}^\nu - \bar{m}^\mu m^\nu = g_0^{\mu\nu} - Vl^\mu l^\nu, \label{gInverse}
\ee
which reproduces the inverse Kerr-Schild metric of equation \eqref{eq:metricKS}.
As discussed above, we will assume that $l^\mu$ is tangent to an SNGC.
Since $l^\mu$ is tangent to a geodesic congruence, it must satisfy
\be
Dl^\mu = fl^\mu . \label{geodesic}
\ee
If the geodesic congruence to which $l^\mu$ is tangent is affinely parameterized, then the function $f$ on the right hand side of \eqref{geodesic} is zero:
\be
Dl^\mu = 0\, .
\ee
Henceforth, we will always assume that geodesic congruences are affinely parameterized unless we explicitly state otherwise. 

\subsection{Optical Scalars} 
\label{sec:optic}

Given a \emph{null} geodesic congruence with tangent vector $l^\mu$, the complex \emph{optical scalars} $\rho$ and $\sigma$, as in e.g. \cite{Adamo:2009vu}, are defined via 
\bea
\rho &:=& m^\nu \bar{m}^\mu \nabla_\mu l_\nu
= m^\nu \bar{\delta} l_\nu\,, \\ \sigma &:=& m^\nu m^\mu \nabla_\mu l_\nu = m^\nu\delta l_{\nu}\,. \label{optical}
\eea
The optical scalars $\rho$ and $\sigma$ are often called the complex divergence and the complex shear. A null geodesic congruence with vanishing $\sigma$ is said to be shear-free. It is common to decompose $\rho = -\theta + i\omega$, where $\theta$ is the expansion scalar, and $\omega$ is the twist. Null geodesic congruences with $\theta \neq 0$, $\omega \neq 0$ are said to be expanding, twisting respectively. 

\begin{figure}[t]
\begin{center}
\includegraphics[width=0.95\textwidth]{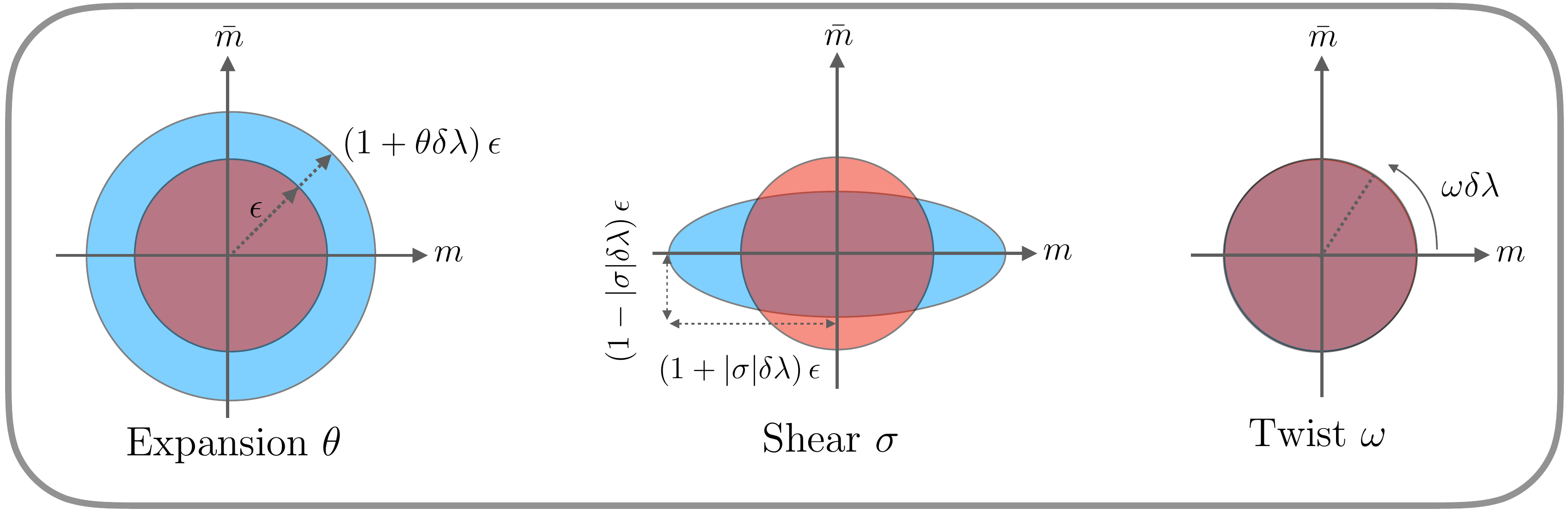}
\caption{ 
Geometric interpretations of the optical scalars. The cross section of a geodesic congruence with non-zero expansion, shear, and twist, is depicted from left to right; assuming a circular cross section (red), the cross section of the congruence a small affine parameter distance $\delta \lambda$ away is depicted (blue).}
\label{fig:OpticalScalars}
\end{center}
\end{figure}

The expansion and twist have the following geometric interpretations. An infinitesimal circle in the plane spanned by $m$ and $\bar{m}$ determines an infinitesimal tube by transporting points on the initial circle a small affine parameter distance $\delta\lambda$ along the null geodesics of the preferred shear-free congruence. When $\sigma = 0$, the final boundary of the tube will also be a circle. If the initial circle has radius $\epsilon$, the final circle will have radius $(1+\theta\delta\lambda)\epsilon$, hence, $\theta$ determines the rate of expansion of the geodesics. Similarly, the final circle will be rotated by an angle $\omega\delta\lambda$, so that $\omega$ measures the degree to which geodesics in the congruence locally twist around each other. When $\rho = 0$, it can be shown that the final boundary of the infinitesimal tube, whose initial boundary was a circle of radius $\epsilon$, will be an ellipse having the same area as the circle,  with semi-major and semi-minor axes given by $(1 \pm |\sigma|\delta\lambda)\epsilon$, while the argument of $\sigma$ determines the orientation of the major axis. A nice discussion of this geometry can be found in chapter 7 of \cite{Huggett:1986fs}. The geometric interpretation of these optical scalars is summarized in figure \ref{fig:OpticalScalars}. 

As defined in \eqref{optical}, the optical scalars $\rho$ and $\sigma$ are two of the spin coefficients of the null tetrad, and it is not immediately obvious that they encode invariant geometrical information about the null congruence generated by $l$ independently of the choice of tetrad. One can however check that when $l$ is tangent to a geodesic, so that the spin coefficient $\kappa$ (discussed in the following section) vanishes, the tetrad transformations belonging to the little group of $l$ leave $\rho$ invariant and can only transform $\sigma$ by an overall phase (see chapter 1 of \cite{Chandrasekhar:1985kt}, for instance). Alternatively, one can verify that (see chapter 6 of \cite{Stephani:2003tm}) 
\be
\theta = \tfrac{1}{2}\nabla_\mu l^\mu, \quad \omega^2 = \tfrac{1}{2}\nabla_{[\mu}l_{\nu]}\nabla^\mu l^\nu, \quad \sigma\bar{\sigma} = \tfrac{1}{2}\nabla_{(\mu}l_{\nu)}\nabla^\mu l^\nu - \theta^2, \label{opticalsFromK}
\ee
which clearly depend only on $l$. We note again for emphasis that for Kerr-Schild spacetimes, the optical scalars are the same whether computed with respect to the background flat metric, or the full spacetime metric.

\subsection{Shear-Free and Geodesic Conditions}

The metric \eqref{full} can be used to describe any expanding Kerr-Schild spacetime, i.e. any Kerr-Schild spacetime admitting an SNGC with $\theta \neq 0$. In the language of the Newman-Penrose formalism, the condition that $l^\mu$ is shear-free is simply $\sigma = 0$, while the geodesic condition becomes $\kappa = 0$, where the spin coefficient $\kappa$ is defined by
\be
\kappa := m^\nu l^\mu \nabla_\mu l_\nu = m^\nu D l_\nu\,. 
\ee
For the null tetrad \eqref{tetrad vectors}, one can directly compute $\kappa$ and $\sigma$ in terms of $\Phi$:
\bea
\kappa &=& \bar{\Phi}_{,\zeta} - \bar{\Phi}\bar{\Phi}_{,u} \,,\label{Sigma}\\ 
\sigma &=& \Phi\left(\bar{\Phi}\bar{\Phi}_{,u} -\bar{\Phi}_{,\zeta}  \right) - \bar{\Phi}\bar{\Phi}_{,\bar{\zeta}} + \bar{\Phi}_{,v}\,. \label{kappa}
\eea
From the expressions \eqref{Sigma} and \eqref{kappa}, it can be seen that $\kappa = 0 = \sigma$ holds whenever $\Phi$ satisfies 
\begin{equation}
D\Phi = 0 = \delta\Phi\,. \label{left leaf}
\end{equation}
 The two conditions \eqref{left leaf} directly imply the following partial differential equations for $\Phi$:
\begin{equation}
\Phi_{,v} = \Phi\Phi_{,\zeta}\,, \qquad \Phi_{,\bar{\zeta}} = \Phi\Phi_{,u}\label{PDEs}\,.
\end{equation}
We note that the general solution to these two equations for $\Phi$ is given by the arbitrary 
analytic function $F$ in the statement of Kerr's theorem,  (\ref{Eqn:KerrTheorem}). Inserting the complex conjugate of \eqref{PDEs} into \eqref{Sigma} and \eqref{kappa} then gives $\kappa = 0 = \sigma$. Differentiating the first differential equation of \eqref{PDEs} with respect to $u$, and the second with respect to $\zeta$, one gets $\Phi_{,uv} = \Phi_{,\zeta\bar{\zeta}}$, which implies that 
\begin{equation}
\Box_0\Phi = 0\, , \label{harmonic}
\end{equation}
so that $\Phi$ is harmonic with respect to the flat-space Laplacian $\Box_0 = 2(\partial_u\partial_v - \partial_\zeta\partial_{\bar{\zeta}})$. Note that $\Phi$ does not satisfy the Pleba\'{n}ski equation  \eqref{eq:PlebEq}. 

The function $\Phi$ can be found in this way for any expanding Kerr-Schild spacetime, including non-stationary and non-vacuum solutions, and will provide the basis for the Newman-Penrose map described in the next section.

\section{The Newman-Penrose Map}
\label{sec:NPDC}

In the previous section we laid the groundwork for understanding Kerr-Schild spacetimes with an expanding SNGC in the Newman-Penrose formalism. We are now in a position to present the main result of this article: given any Kerr-Schild spacetime, whose Kerr-Schild vector $l^\mu$ is tangent to an expanding SNGC, one may identify a certain complex harmonic function \eqref{harmonic} that enjoys a distinguished role in the construction of a preferred null tetrad aligned with the tangent vector $l^\mu$ of the SNGC \eqref{eq:tetrad}. One can then associate to any such spacetime a self-dual solution of the vacuum flat-space Maxwell equations defined by the Newman-Penrose map  $A = \hat{k}\Phi$.

While this construction is, a priori, independent of previous approaches to the classical double copy, we argue that it is in fact a novel manifestation of the same correspondence. We provide evidence for this claim by showing that the real part of the self-dual gauge fields associated to the Schwarzschild and Kerr spacetimes by the Newman-Penrose double copy are gauge equivalent to the usual single copy gauge fields obtained from the standard Kerr-Schild double copy outlined in \ref{KSDC}. We also examine the Kinnersley photon rocket solution \cite{Kinnersley:1969zz}, which we find under our correspondence to be gauge equivalent to the Li\'{e}nard-Wiechert potential of a moving point charge, which is a slightly different single copy than found in previous work \cite{Luna:2016due}.

\subsection{General Procedure}\label{subsec:generalProcedure}
It was shown in \cite{Monteiro:2014cda, Berman:2018hwd} that, up to a constant scale factor which we have chosen on dimensional grounds, the Kerr-Schild operator $\hat{k}$ of the self-dual double copy  takes the form 
\begin{equation}
\hat{k} = -\frac{Q}{2\pi\epsilon_0}(dv\,\partial_\zeta + d\bar{\zeta}\,\partial_u)\,, \label{diffOp}
\end{equation}
where $Q$ is the total electric charge, and $\epsilon_0$ is the vacuum permittivity. 
The form of \eqref{diffOp} can be  naturally understood in terms of the spinorial formalism,  the details of which are given in \ref{app:khat}.

It is now a simple matter to show that $A := \hat{k}\Phi$ is self-dual, and therefore solves the vacuum Maxwell equations. Indeed, from \eqref{diffOp} we have 
\begin{equation}
A = -\frac{Q}{2\pi\epsilon_0}\left(\Phi_{,\zeta}dv + \Phi_{,u}d\bar{\zeta}\right), \label{gauge field}
\end{equation}
from which we can compute the field strength two-form $F = dA$: 
\begin{eqnarray}
F = -\frac{Q}{2\pi\epsilon_0}\left(\Phi_{,u\zeta}du\wedge dv -\Phi_{,\zeta\zeta}dv\wedge d\zeta + \Phi_{,uu}du\wedge d\bar{\zeta} + \Phi_{,u\zeta}d\zeta\wedge d\bar{\zeta}\right) ,
\label{SD field strength}
\end{eqnarray}
where we used $\Phi_{,uv} - \Phi_{,\zeta\bar{\zeta}} = \tfrac{1}{2}\Box_0\Phi = 0$. The field strength in \eqref{SD field strength} is self-dual with respect to $\star_0$, the Hodge star operator associated with the background metric $g_0$---that is, $F$ satisfies
\begin{equation}
F =  i\star_0 F\,, \label{anti self-dual}
\end{equation}
which, together with the fact that $F$ is exact, implies the vacuum flat-space Maxwell equation 
\begin{equation}
d \star_0 \! F = 0\,.
\end{equation}
The above considerations demonstrate that given any flat-space harmonic function $\Phi$, one can define a self-dual solution of the vacuum flat-space Maxwell equations by $A = \hat{k}\Phi$. On the other hand, in the previous section we showed that every Kerr-Schild spacetime with an expanding SNGC has associated with it a flat-space harmonic function $\Phi$. It is then clear that every Kerr-Schild spacetime with an expanding SNGC can be mapped to a self-dual solution of the vacuum flat-space Maxwell equations. We call this correspondence \emph{the Newman-Penrose map}, which we summarize as follows:  
\begin{enumerate}
     \item{Consider a Kerr-Schild spacetime whose Kerr-Schild vector $l^\mu$ is tangent to an expanding SNGC.}
    \item {By Kerr's Theorem, a harmonic function $\Phi$ will arise in the construction of a preferred null tetrad aligned with $l^\mu$.}
    \item{We can then construct a gauge field $A = \hat{k} \Phi$, which is automatically a self-dual solution of the vacuum Maxwell equations when $\hat{k}$ is given by \eqref{diffOp}.}
\end{enumerate}

 We now proceed to study some concrete examples of the Newman-Penrose map which illustrate its close relationship to the Kerr-Schild and self-dual double copies summarized in \ref{sec:clDC}.

\subsection{Examples}
\label{sec:examples}
The Newman-Penrose map defined in the previous subsection is, a priori, independent of the usual Kerr-Schild double copy which we reviewed in \ref{KSDC}, and one would not necessarily expect there to be any clear relationship between the real gauge fields associated with each prescription. We do not presently have a proof of a general relationship between the Kerr-Schild double copy and Newman-Penrose map, nevertheless, we are encouraged to believe that there may be a deep connection between the two. Below we study the Newman-Penrose map applied to examples of vacuum black hole and pure radiation solutions, where we find that the gauge field associated to the spacetime by the Newman-Penrose map and the gauge field associated to the same spacetime by the Kerr-Schild double copy agree exactly in the case of vacuum solutions, and agree up to some subtleties discussed below in the case of a pure radiation solution. 

\subsubsection{The Schwarzschild Black Hole}
\label{eq:exampleSchwarzschild}
The tetrad \eqref{eq:tetrad} associated with the Schwarzschild metric is presented in  \cite{mcintosh1988single}. The complex scalar $\Phi$ is given by
\begin{equation}
\Phi = \frac{1}{2\bar{\zeta}}\left(v-u- \sqrt{2} r\right),
\label{Schwarzschild zero copy}
\end{equation}
where $r= \sqrt{x^2+y^2+z^2} = \frac{1}{\sqrt{2}}\sqrt{(v-u)^2 + 4 \zeta \bar{\zeta}} $ is the usual radial coordinate written in light-cone coordinates \eqref{eq:coor}. In \ref{app:ConfirmMetric}, we will show in detail that this form for $\Phi$ generates the correct metric.

For any vacuum Kerr-Schild spacetime, the real function $V$ is given by (see for instance \cite{Stephani:2003tm})
\be
V = \frac{M(\rho + \bar{\rho})}{2P^3} \,,
\label{V-Schwarzschild}
\ee
where $\rho$ is the complex expansion defined in section \ref{sec:optic}, and $P = a + b \, \Phi + \bar{b} \, \bar{\Phi} + c \, \Phi\bar{\Phi}$, with $a,c$ real constants and $b$ a complex constant. For the Schwarzschild metric, we have
\be
P = \frac{1}{\sqrt{2}}(1 + \Phi\bar{\Phi}) , \qquad \rho = \frac{1 + \Phi\bar{\Phi}}{2\Phi\bar{\zeta} - (v-u)}\label{P+rho}\,.
\ee
Equations \eqref{Schwarzschild zero copy}, \eqref{V-Schwarzschild} and \eqref{P+rho} determine the null tetrad \eqref{eq:tetrad}, which in turn determines the metric \eqref{full}. 
The invariant line element in Cartesian coordinates is given by  
\be
ds^2 = dt^2 - dx^2 - dy^2 - dz^2 - \frac{2GM}{r}(dt + dr)^2, \label{Schwarzschild Cartesian}
\ee
 which is the Kerr-Schild form of the Schwarzschild metric (see \ref{app:ConfirmMetric} for details). Transforming the time coordinate $t = T + 2GM\ln(\tfrac{r}{2GM}-1)$, and using the usual spherical coordinates, one recovers the more familiar form  of the metric in Schwarzschild coordinates
\be
ds^2 = \left(1-\frac{2GM}{r}\right)dT^2 - \left(1-\frac{2GM}{r}\right)^{-1}dr^2 - r^2 d\Omega^2.
\ee
From \eqref{Schwarzschild Cartesian} we can read off the Kerr-Schild single copy gauge field
\be
A_{\mbox{\tiny KS}} = \frac{2GM}{r}(dt + dr) \,, \label{KS couloumb}
\ee
which is gauge equivalent to the standard Coulomb field 
\be
\label{couloumb}
A_{\mbox{\tiny Coulomb}} = \frac{Q}{4\pi\epsilon_0 r}dt\,,
\ee
where we have made the replacements $8\pi G \to 1/\epsilon_0$, and $M \to Q$. 

Using equation \eqref{Schwarzschild zero copy}, we can now compute the Newman-Penrose single copy gauge field via $A_{\mbox{\tiny NP}} = \hat{k}\Phi  \nonumber$:
\begin{eqnarray}
A_{\mbox{\tiny NP}} = {\frac {Q}{2\sqrt{2} \pi\epsilon_0r }}\,\left(dv -\Phi d\bar{\zeta} \right).
\label{Schwarzschild complex single copy}
\end{eqnarray}
Transforming to spherical coordinates via
\begin{equation}
 \theta = \tan^{-1}\left(\frac{\sqrt{x^2 + y^2}}{z}\right), \qquad \phi = \tan^{-1}\left(\frac{y}{x}\right),
\end{equation}
one finds 
\begin{align}
\label{eq:SchA}
A_{\mbox{\tiny NP}} = \frac{Q}{4\pi\epsilon_0} \left( \frac{dt}{r} - i (1-\cos\theta)d\phi +  \frac{dr}{r}-\frac{\sin\theta}{(1+\cos\theta)} d\theta\right).
\end{align}
The last two terms in $A_{\mbox{\tiny NP}}$ are exact, so the real part of $A_{\mbox{\tiny NP}}$ can be written 
\begin{equation}
\mathfrak{Re}(A_{\mbox{\tiny NP}}) = \frac{Q}{4\pi\epsilon_0 r}dt + A_{\mbox{\tiny gauge}}\,,
\end{equation}   
with a pure gauge contribution of the form 
\begin{eqnarray}
A_{\mbox{\tiny gauge}} =  \frac{Q}{4\pi\epsilon_0} d\ln\bigl( r(1+\cos\theta)\bigr)\,,
\label{eq:Schgauge}
\end{eqnarray}
in perfect agreement with \eqref{couloumb}. On the other hand, the imaginary part of $A_{\mbox{\tiny NP}}$ is 
\begin{equation}
\mathfrak{Im}(A) =- \frac{Q}{4\pi\epsilon_0}\left(1- \cos\theta\right)d\phi\,,
\end{equation}
which is the vector potential for a magnetic monopole, defined away from the negative $z$-axis. The magnetic monopole defined away from the positive $z$-axis i.e. $\mathfrak{Im}(A) \propto (1 + \cos \theta) d \phi$, 
arises by repeating this procedure with a parity transformation $\vec{x} \rightarrow - \vec{x}$ in the usual way \cite{Coleman:1982cx}. Note that
$\mathfrak{Im}(A)$ 
has a singularity at $\theta= \pi$, which is not surprising 
since that is the location of the Dirac string. However, it is surprising that 
the pure gauge part appearing in \eqref{eq:Schgauge} for $\mathfrak{Re}(A)$ is singular at $\theta = \pi$, since in principle one can have different covers for $\mathfrak{Re}(A)$ 
and $\mathfrak{Im}(A)$ while still preserving the self-dual condition for $F$. So it is not {\em a priori} obvious that the singularities of $\mathfrak{Re}(A)$ 
and $\mathfrak{Im}(A)$ should be related in any way. We will return to this observation in future work.

\subsubsection{The Kerr Black Hole}
\label{subsec:KerrBH}
We now turn to the more general case of the Kerr spacetime. The adapted null tetrad for Kerr is given in \cite{mcintosh1988single}. In particular, the complex harmonic scalar for Kerr is obtained from the corresponding function in the null tetrad for Schwarzschild by applying the Newman-Janis trick \cite{mcintosh1988single, Erbin:2016lzq}. Namely, one performs a complex coordinate transformation of the form
\begin{equation}
z \to z + ia \,,
\end{equation}
or equivalently in terms of light-cone coordinates
\begin{eqnarray}
v - u &\to& v - u + \sqrt{2}ia \,, \label{NJtrick}
\end{eqnarray}
where $a = J/M$ is the angular momentum per unit mass of the Kerr black hole.
After performing the Newman-Janis trick \eqref{NJtrick} on the Schwarzschild complex scalar \eqref{Schwarzschild zero copy}, we obtain the complex scalar of the null tetrad for the Kerr spacetime
\bea
\nonumber
\Phi^{\mbox{\tiny (Kerr)}} &=& \frac{1}{2\bar{\zeta}}\left((v - u + \sqrt{2}ia)-\sqrt{(v - u + \sqrt{2}ia)^2 + 4\zeta\bar{\zeta}}\right) \\ 
&=&\frac{r-ia}{2\bar{\zeta}r}\left(v - u -\sqrt{2} r\right),
\label{eq:PhiKerr}
\eea
which by a direct computation can be seen to satisfy \eqref{PDEs}. See \ref{app:ConfirmMetric} for a detailed verification that the form of $\Phi^{\mbox{\tiny (Kerr)}}$ in \eqref{eq:PhiKerr} generates the correct metric. Here $r$ is defined implicitly by
\begin{align}
\frac{x^2+y^2}{r^2+a^2} + \frac{z^2}{r^2} = 1\label{rdef}\,,
\end{align}
so that in light-cone coordinates, making use of $z =r \cos\theta =  (v-u) / \sqrt{2}$, this becomes
\begin{align}
\sqrt{2}r+\frac{ia}{r}(v-u) &= \sqrt{(v-u+\sqrt{2}i a)^2 + 4 \zeta \bar{\zeta}}  \,.
\end{align}
We have the same equations as the Schwarzschild case for the functions $V$ \eqref{V-Schwarzschild}, $P$ and $\rho$ \eqref{P+rho}, with $\Phi$ replaced with $\Phi^{\mbox{\tiny (Kerr)}}$ and $v - u$ replaced with $v - u + \sqrt{2}ia$ after the Newman-Janis trick.

The Kerr metric in Kerr-Schild form is given by  \eqref{eq:metricKS} with
\begin{align}
\label{eq:KerrKS}
\varphi = \frac{2GM r^3}{r^4+a^2 z^2}
\,, \quad{}k_\mu dx^\mu &=dt + \frac{rx+ay}{r^2+a^2} dx + \frac{ry-ax}{r^2+a^2}dy + \frac{z}{r} dz\,,
\end{align} written in Cartesian coordinates. The usual Kerr-Schild single copy is
\begin{equation}
A^{\mbox{\tiny (Kerr)}}_{\mbox{\tiny KS}}= \frac{Q r}{4\pi \epsilon_0  (r^2 + a^2 \cos^2\theta)}\left( dt + \frac{rx + ay}{r^2+a^2} dx + \frac{ry-ax}{r^2+a^2} dy + \frac{z}{r} dz\right),
\label{KerrKSGauge}
\end{equation}
again with the replacement $8\pi G \rightarrow 1/\epsilon_0$ and $M\rightarrow Q$. 
This solution to Maxwell's equations, referred to as $\sqrt{\textnormal{Kerr}}$ by \cite{Arkani-Hamed:2019ymq}, can be written in spherical coordinates as
\begin{align}
A_{\sqrt{\textnormal{Kerr}}} = \frac{Q r}{4\pi \epsilon_0  (r^2 + a^2 \cos^2\theta)}\left( dt +dr - a\sin^2\theta d\phi\right),
\end{align}
with the coordinate transformations
\begin{align}
\label{eq:sph}
x+iy = (r+ia)e^{i\phi} \sin\theta  \quad \text{and} \quad
z = r\cos\theta \,.
\end{align}
This has been shown \cite{Monteiro:2014cda,Arkani-Hamed:2019ymq} to describe the field produced by an axisymmetric charge distribution rotating at a uniform rate about the $z$-axis.

For the Newman-Penrose map, we can now apply the operator $\hat{k}$ of \eqref{diffOp} to $\Phi^{\mbox{\tiny (Kerr)}}$ \eqref{eq:PhiKerr} to obtain the self-dual gauge field
\begin{eqnarray}
A ^{\mbox{\tiny (Kerr)}}_{\mbox{\tiny NP}}
= {\frac {Q}{2\sqrt{2} \pi\epsilon_0(r+ia\cos\theta)}}\,\left(dv- \Phi^{\mbox{\tiny (Kerr)}} d\bar{\zeta}\right).
\end{eqnarray}
Equivalently, this gauge field could have been obtained 
directly from (\ref{Schwarzschild complex single copy}) after applying the Newman-Janis trick. Transforming to spherical coordinates via \eqref{eq:sph} gives
\begin{align}
A ^{\mbox{\tiny (Kerr)}}_{\mbox{\tiny NP}} =& \frac{Q}{4\pi \epsilon_0}\bigg[\frac{r}{r^2+a^2\cos^2\theta}\left(dt + dr - a\sin^2\theta d\phi\right)-\frac{\sin\theta d\theta}{1+\cos\theta}\nonumber\\
&\ \ \ \ \ \ \ \ -\frac{ia\cos\theta}{(r^2+a^2\cos^2\theta)}(dt + dr) - i(1-\cos\theta)\frac{r^2-a^2\cos\theta}{r^2+a^2\cos^2\theta}d\phi\bigg].
\label{KerrComplexGaugeField}
\end{align}
Taking the real part of \eqref{KerrComplexGaugeField} leads to 
\begin{align}
\mathfrak{Re}(A^{\mbox{\tiny (Kerr)}}_{\mbox{\tiny NP}}) =&\ A^{\mbox{\tiny (Kerr)}}_{\mbox{\tiny gauge}}+ \frac{Qr}{4\pi \epsilon_0(r^2+a^2\cos^2\theta)}\left(dt + dr - a\sin^2\theta d\phi\right),
\end{align}
where
\be
A^{\mbox{\tiny (Kerr)}}_{\mbox{\tiny gauge}} = \frac{Q}{4\pi\epsilon_0} d\ln\left(1+\cos\theta\right).
\ee
This solution agrees exactly with $\sqrt{\textnormal{Kerr}}$ in \eqref{KerrKSGauge}.

 In accordance with electromagnetic duality \cite{figueroa1998electromagnetic}, we expect that the imaginary part of the field,
 \begin{align}
\mathfrak{Im}(A ^{\mbox{\tiny (Kerr)}}_{\mbox{\tiny NP}}) =& -\frac{Q}{4\pi \epsilon_0}\bigg[\frac{ a\cos\theta}{r^2+a^2\cos^2\theta} (dt + dr)  +  (1-\cos\theta)\frac{r^2-a^2\cos\theta}{r^2+a^2\cos^2\theta} d\phi\bigg],
\end{align}
is sourced by an axisymmetric distribution of \emph{magnetic} charge rotating at a uniform rate about the $z$-axis. As in the static case, we expect the imaginary part of the field to have support on a subset of spacetime with non-trivial topology, but we will not digress into a detailed analysis of these features here. 

We conclude this example with a na\"ive discrepancy and a puzzle. 
The Kerr-Schild form for the Kerr metric obtained directly using $\Phi$ is
\begin{align} 
ds^2&=ds^2_0 - \frac{2 M r}{(r^2+a^2\cos^2\theta)}\left(dt -\frac{z}{r} dz- \frac{rx+ay}{(r^2+a^2)}dx-\frac{ry-ax}{(r^2+a^2)}dy\right)^2,
\label{eqn:Kerr-from-Phi}
\end{align}
which we demonstrate in \ref{app:ConfirmMetricSubSec}. This expression 
differs from the double copy form for the Kerr metric \eqref{eq:KerrKS}. 
In \ref{app:Kerr-metric-many-forms} we show however, that these two expressions for 
the metric are equivalent, being related by a coordinate 
transformation, so there is no contradiction. Briefly, starting from 
standard textbook expressions for the Kerr metric, such as the Boyer-Lindquist form, 
one arrives at these two different Kerr-Schild metrics by following 
either the ingoing or outgoing principal null direction ray. What is surprising 
is that depending on the single copy prescription---either the usual one 
or Newman-Penrose one presented here---the same gauge equvialent potential $A_{\sqrt{\rm Kerr}}$ (at least for the real part) arises from two different 
Kerr-Schild metrics that are related by a coordinate change. 
The same feature also occurs in 
the Schwarzschild case. This result suggests that the 
Newman-Penrose map may provide further insights into the relationship between 
the gauge symmetries on both sides of the correspondence.

 \subsubsection{Kinnersley's Photon Rocket} \label{Sec:PhotRock}
For our final example, we study an exact solution of Einstein's equations sourced by a pure radiation stress energy tensor of the form
\be
T_{\mu\nu} = f k_\mu k_\nu\,, \label{stress}
\ee
known as Kinnersley's photon rocket \cite{Kinnersley:1969zz}, which describes the gravitational field produced by a particle moving along an arbitrary timelike worldline. 

\begin{figure}[t]
\begin{center}
\includegraphics[width=0.35\textwidth]{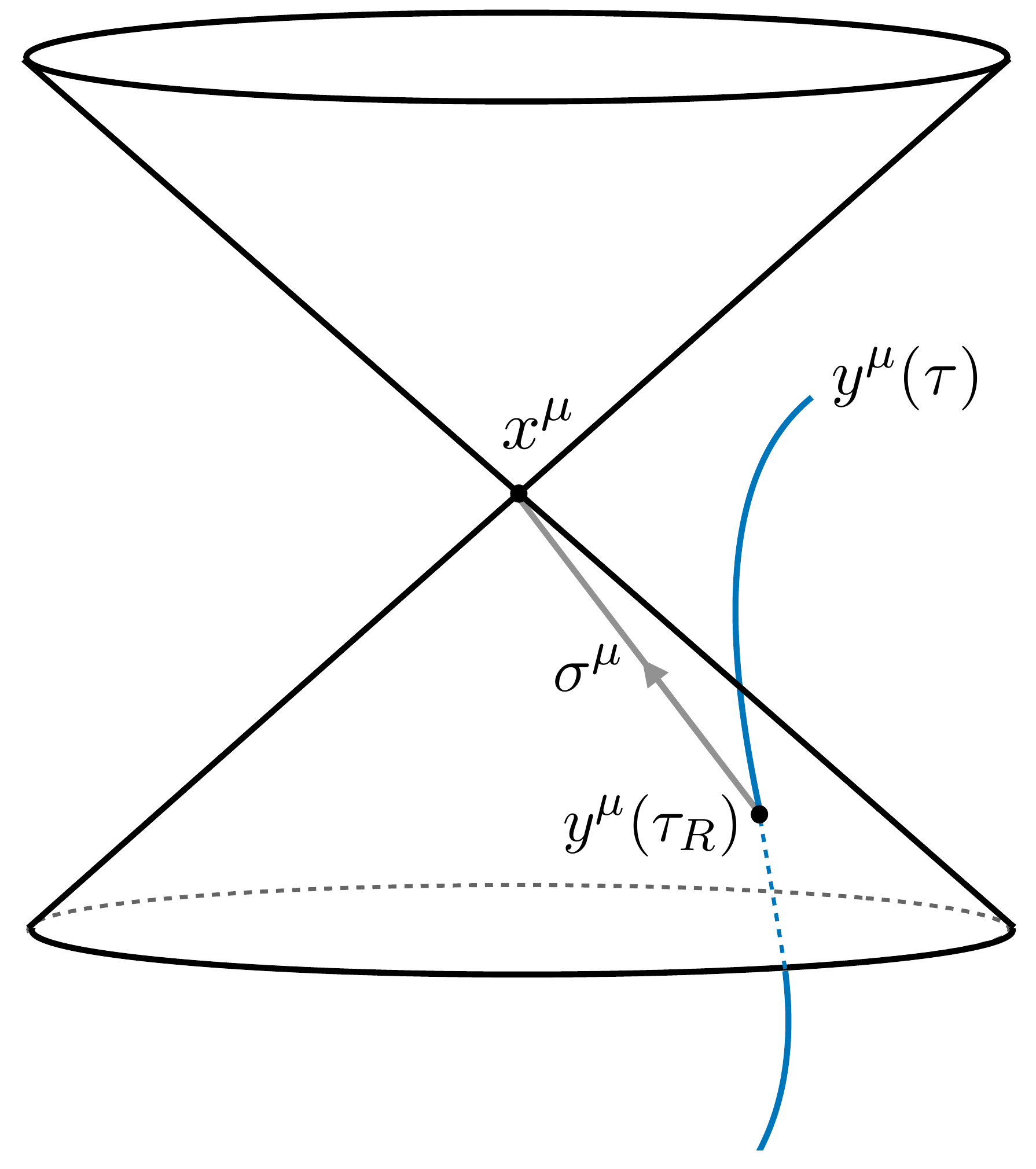}
\caption{An arbitrary timelike worldline $y^\mu(\tau)$, parameterized with respect to proper time $\tau$, intersects the past light-cone of an arbitrary point $x^\mu$ in Minkowski space at exactly one point $y^\mu(\tau_R)$. This picture serves to define of $\tau_R(x)$. }
\label{fig:photrock}
\end{center}
\end{figure}

To describe the particle's motion, we define $y^\mu(\tau)$ as the coordinates of the particle moving along a worldline parameterized by proper time $\tau$. The worldline of the particle will intersect the past light-cone of a point $x^\mu$ at exactly one point $y^\mu(\tau_R)$, which implicitly defines the retarded proper time $\tau_R$ (see figure \ref{fig:photrock}).

The retarded distance $r$ is defined by
\begin{align}
r = \sigma \cdot \lambda(\tau_R)
\end{align}
where $\sigma^\mu = x^\mu - y^\mu(\tau_R)$, and  $\lambda^\mu(\tau) = \tfrac{d y^\mu(\tau)}{d\tau}$ 
is the proper velocity of the particle normalized such that $\lambda\cdot \lambda = 1$. Here indices are raised and lowered with respect to the background metric $g_0$. From a geometric standpoint, the proper distance $r$ may be regarded as an affine parameter along the future-pointing null geodesics originating on the worldline $y^\mu(\tau)$. For more details on the geometric setup see \cite{Carlip:1999an, Bonnor:1994ir, newman1963class}. 

In Kerr-Schild form, the metric for the photon rocket can be written as in  \eqref{eq:metricKS} with 
\begin{align}
\varphi = \frac{2GM(\tau_R)}{r} \quad \text{and} \quad
k_\mu = \frac{\sigma_\mu}{r}\,.
\end{align}
In \cite{Luna:2016due}, this solution (with $M(\tau_R) = M =$ constant) was analyzed in the context of the double copy, where it was shown to
correspond to a gauge field of the form
\begin{align}
A^{\mbox{\tiny (PR)}}_{\mbox{\tiny KS}}  = \frac{Q}{4\pi \epsilon_0 r^2}\sigma_\mu dx^\mu,
\end{align}
which is the potential of a boosted point charge, but does not contain the radiation field necessarily produced by an accelerating charged particle \cite{Carlip:1999an}. This solution does not satisfy the vacuum Maxwell equations, but instead has a source of the form
\begin{align}
\partial_\mu F^{\mu\nu} = j^\nu \quad \text{with} \quad
j^\nu = \frac{Q}{2\pi \epsilon_0 r^2} (\dot{\lambda}(\tau_R)\cdot k) k^\nu, \label{current}
\end{align}
where overdots denote differentiation with respect to $\tau_R$. Comparing \eqref{stress} and \eqref{current}, we see that in this case, the Kerr-Schild double copy ansatz leads to a double copy between the sources of the electromagnetic and gravitational fields.

The Kerr-Schild vector of the photon rocket spacetime is expanding and shear-free, so we can apply the Newman-Penrose map to the scalar $\Phi^{\mbox{\tiny (PR)}}$
which satisfies \eqref{PDEs} and generates a self-dual gauge field. Using the fact that $\sigma_\mu$ lies on the lightcone, we can rewrite in lightcone coordinates
\begin{align}
\nonumber
 (x - y(\tau_R))\cdot (x - y(\tau_R)) &= 0 \\
\Rightarrow \,\,\, (\zeta - \zeta_0)(\bar{\zeta}- \bar{\zeta}_0) &= (u - u_0)(v - v_0) , \label{cone}
\end{align}
with $y^\mu (\tau_R) \equiv (u_0,\, v_0,\,\zeta_0,\, \bar{\zeta}_0)$ and $u_0  \equiv u_0(\tau_R)$ etc. This equation implicitly defines $\tau_R$ in terms 
of the coordinates $x^\mu$, and extends $\sigma$ to a function of $x^\mu$. Taking the exterior derivative and solving for $d\tau_R$, we find
\begin{align}
\nonumber
\label{eq:dtau}
d\tau_R
&= \frac{1}{r}\left[(v-v_0) du+(u-u_0) dv   - (\zeta - \zeta_0) d\bar{\zeta} - (\bar{\zeta} - \bar{\zeta}_0) d\zeta\right]\\
&= \frac{1}{r} \sigma_\mu dx^\mu,
\end{align}
which is exactly the Kerr-Schild vector. Comparing with \eqref{eq:tetrad}, we see that in this case the complex harmonic scalar for the photon rocket takes the form
\begin{align}
\Phi^{\mbox{\tiny (PR)}} = -\frac{(\zeta - \zeta_0)}{(v - v_0)}\,.
\end{align}
 Applying the operator $\hat{k}$ and taking the real part yields 
\begin{align}
\mathfrak{Re}\left(A_{\mbox{\tiny NP}}^{\mbox{\tiny (PR)}}\right) &= \mathfrak{Re}\left(A_{\mbox{\tiny gauge}}^{\mbox{\tiny (PR)}}\right) + \frac{Q}{4\pi \epsilon_0 r} \left(\dot{u}_0 dv + \dot{v}_0 du - \dot{\zeta}_0 d\bar{\zeta} - \dot{\bar{\zeta}}_0 d\zeta\right),
\end{align}
where
\begin{align}
\mathfrak{Re}\left(A_{\mbox{\tiny gauge}}^{\mbox{\tiny (PR)}}\right) = \frac{Q}{4\pi \epsilon_0 }d\ln\left(v - v_0\right).
\label{eq:PRgauge}
\end{align}
Without the pure gauge term, this can be rewritten in the form 
\begin{align}
\mathfrak{Re}\left(A_{\mbox{\tiny NP}}^{\mbox{\tiny (PR)}}\right) &=  \frac{Q}{4\pi \epsilon_0 r} \lambda_\mu (\tau_R)dx^\mu,
\end{align}
which is a Li\'{e}nard-Wiechert potential, i.e., the field produced by a point charge with arbitrary acceleration \cite{Jackson:1998nia}. 
As with the fields, there is no explicit double copy structure for the source terms ($T_{\mu\nu} \propto k_\mu k_\nu$ vs. $\partial_\mu F^{\mu\nu} = 0$) for the Newman-Penrose map. However, our prescription leads to a more intuitive and physical gauge field than the one produced by the Kerr-Schild double copy \cite{Luna:2016due}. In fact, the source term on the gauge theory side found by \cite{Luna:2016due} was added in order to obtain exactly the gauge field that we instead obtain directly from our straightforward prescription.

The imaginary part of $A_{\mbox{\tiny NP}}^{\mbox{\tiny (PR)}}$ is 
\begin{align}
\mathfrak{Im}(A_{\mbox{\tiny NP}}^{\mbox{\tiny (PR)}}) = \mathfrak{Im}(A_{\mbox{\tiny gauge}}^{\mbox{\tiny (PR)}}) - \frac{iQ}{4\pi \epsilon_0 r}\bigg[\frac{\zeta - \zeta_0}{v-v_0}(\dot{v}_0 d\bar{\zeta} - \dot{\bar{\zeta}}_0 dv) + \frac{\bar{\zeta}-\bar{\zeta}_0}{u - u_0}(\dot{u}_0 d\zeta - \dot{\zeta}_0 du)\bigg]
\end{align}
with the pure gauge term
\vspace{-0.08 in}
\begin{align}
\mathfrak{Im}(A_{\mbox{\tiny gauge}}^{\mbox{\tiny (PR)}}) = \frac{iQ}{4\pi \epsilon_0} d\ln(\zeta- \zeta_0)
\label{Im-gauge}
\end{align}
The singularities appearing in the real and imaginary pure gauge terms include the expected singularity at $r = 0$, the position of the worldline of the particle.


\subsection{The Many-to-One Nature of the Map}
\label{subsec:ManyToOne}
General relativity is far richer than electromagnetism, and as such it is remarkable that a classical double copy prescription yields a map between the two, even for a restricted class of solutions.  
However, because of the fundamental differences between the theories, not all components of each theory should be expected to have well-defined analogs on either side of the map. Indeed, such ambiguities have hindered progress in understanding the double copy in the context of charged black holes and certain classes of radiating solutions. A novel aspect of the Newman-Penrose map is its treatment of such ambiguities, and in particular we comment in this subsection on the way in which components that have no obvious gauge theory analog are ``projected out".

The Einstein equations admit black hole solutions, both static (Reissner-N\"{o}rdstrom) and spinning (Kerr-Newman), which carry electromagnetic charge i.e. $T_{\mu \nu}$ is the Maxwell stress energy tensor. When electric charge is included in the gravitational theory,  there is an ambiguity in how charge on the gravitational side would map to a gauge theory quantity under the double copy; indeed, there is very little literature dealing with such solutions.\footnote{\footnotesize{Under the prescription introduced in \cite{Bah:2019sda}, the  stress energy tensor itself may be put into a form of an electromagnetic current.}} 
These charged black holes are algebraically special spacetimes of Petrov type D admitting an SNGC, and may be put into Kerr-Schild form (see \cite{Stephani:2003tm} for a review). Therefore, the Newman-Penrose map may be applied to charged black hole solutions, and the preferred null tetrad is given by \eqref{eq:tetrad} with 
\begin{align}
V \propto  \frac{M}{r} - \frac{Q_{\rm em}^2}{r^2}\,,
\end{align}
where $Q_{\rm em}$ is the electromagnetic charge of the black hole, in contrast with $V\propto \frac{M}{r}$ for uncharged black holes.
Critically, the electric charge is encoded in $V$, not in the complex scalar $\Phi$, and $\Phi$ will be of exactly the same form as for the uncharged Schwarzschild and Kerr black holes: \eqref{eq:PhiKerr} and \eqref{Schwarzschild zero copy}, respectfully (as must be the case so that the limit $Q_{\rm em} \rightarrow 0$ reproduces these solutions). 
Therefore, the Newman-Penrose map is insensitive to the electromagnetic charge associated with the black hole; the information of the electric charge on the gravitational side of the duality, which has no clear gauge theory counterpart, has been “projected out.” The result of the Newman-Penrose map applied to charged black holes will be identical to the result obtained from applying the map to the corresponding uncharged solution with $Q_{\rm em}$ set to zero.

A similar situation arises when one considers certain pure radiation spacetimes with a Bondi mass aspect specified as a function of a retarded or advanced time. The physical intuition for such spacetimes is clear: they describe the gravitational fields produced by compact massive objects that are absorbing or emitting radiation, say in the form of a null dust. In accordance with mass-energy equivalence, the mass of such an object will vary as it absorbs or emits radiation. However, no such analogous process exists in electrodynamics; a charged object can absorb or emit radiation, but such a process cannot lead to a change in the total charge of the system. A straightforward example is provided by the Newman-Penrose map for the photon rocket of section \ref{Sec:PhotRock}, where the Bondi mass aspect $M(\tau_R)$ appears only in the function $V$ but not in the complex harmonic scalar $\Phi$. In the simple case where the worldline is chosen to be a straight, time-like \emph{line} with vanishing spatial acceleration, e.g., the line determined by $x = y = z = 0$ in Cartesian coordinates, the photon rocket reduces to the Vaidya metric. In this case, it may be observed that the complex harmonic scalar $\Phi$ for the photon rocket reduces to that of the Schwarzschild spacetime \eqref{Schwarzschild zero copy}, while the function $V$ reduces to the function $V$ of the Schwarzschild spacetime with the constant mass $M$ replaced by the Bondi mass aspect $M(\tau_R)$. Whether in the case of the general worldline, or in the zero acceleration limit, the conclusion is the same: the Newman-Penrose map is insensitive to the functional form of the Bondi mass aspect, producing the same self-dual gauge field for any choice of $M(\tau_R)$. As for the case of charged black holes, the Newman-Penrose map ``projects out" the information on the gravitational side that has no sensible analog on the gauge theory side, and maps many physically distinct solutions on the gravity side to the same self-dual gauge field.

In this way the many-to-one nature of the Newman-Penrose map provides a novel way to treat degrees of freedom that have no well-defined analog on the gauge
side of the correspondence: an advantage over the traditional classical double copy in which a prescription for dealing with such ambiguities is not fully understood. 

\section{Discussion}
\label{sec:discussion}
In this work, we have introduced a novel correspondence between Kerr-Schild spacetimes with an expanding SNGC and self-dual solutions of the vacuum Maxwell equations. Since this correspondence makes use of the null tetrad formalism, we have christened it the \emph{Newman-Penrose map}.  We have applied this map to three concrete examples: the Schwarzschild spacetime, the Kerr spacetime, and Kinnersley's photon rocket. In each case, we have shown that the real part of the self-dual gauge field defined by this map connects with the standard results of the classical double copy up to gauge transformations. Additionally, the imaginary part of the self-dual gauge field is sourced by the electromagnetic dual of the real part. We showed this explicitly in the Schwarzschild example for which the imaginary part of the gauge field reproduces a magnetic monopole. In the case of the photon rocket, the Newman-Penrose map seems to provide a genuine advantage: we obtain a field that is gauge-equivalent to the Li\'{e}nard-Wiechert potential without the need to introduce any ad-hoc arguments involving radiation fields as in \cite{Luna:2016due}. 

Our approach has some limitations as well as some notable advantages. Since we use the null tetrad formalism, it is not clear whether or not our results can be readily extended beyond four spacetime dimensions, whereas for the usual Kerr-Schild double copy, such an extension is trivial \cite{Carrillo-Gonzalez:2017iyj,Monteiro:2014cda}. Since our results are limited to Kerr-Schild spacetimes with an expanding SNGC, they cannot be applied to vacuum Kerr-Schild spacetimes of Petrov type N, which have vanishing expansion; in particular, we cannot reproduce results relating PP wave spacetimes to electromagnetic plane wave solutions as in e.g. \cite{Luna:2018dpt}. However, as we have seen in the example of the photon rocket, \emph{our results can be applied directly to Kerr-Schild spacetimes that are neither stationary, nor vacuum}---a significant advantage compared with the usual Kerr-Schild double copy. Furthermore, the Newman-Penrose map projects out information that has no sensible analog on the gauge side of the duality. 

In this work, we have restricted our considerations to the case of real, single  Kerr-Schild metrics, but it would be interesting to investigate whether, as in the case of the Kerr-Schild double copy, our construction can be generalized to include complex or double Kerr-Schild metrics as well. Given the prominent role of self-dual gauge fields and electromagnetic duality in our construction, it would be particularly nice to make contact with \cite{Huang:2019cja, Luna:2015paa,Alawadhi:2019urr} by relating the Taub-NUT metric to a (self-dual) dyon via a suitably generalized Newman-Penrose map. We also hope to extend this formalism to include complex spacetimes, in particular Eguchi-Hanson space studied in the context of the classical double copy previously in \cite{Berman:2018hwd}.

Another potential generalization of the Newman-Penrose map would 
be to non-trivial spacetimes that are asymptotically flat. The remarkable fact that, for any arbitrary asymptotically flat-spacetime, there is a
generalization of 
Kerr's theorem which generates 
null geodesic congruences that are asymptotically shear-free \cite{Newman:2005qn}, hints at such a possibility. 
An interesting topic for further exploration would be whether such asymptotically shear-free null geodesic congruences, and the three-dimensional 
Cauchy-Riemann manifolds they define (see \cite{Adamo:2009vu} and references therein), map to asymptotic solutions
to the flat-space source-free Maxwell's equations.

In addition to discovering new solutions that admit a classical interpretation under the Newman-Penrose map, it would be interesting to explore the ways in which this novel map may offer insights into various  outstanding questions. For instance, while the double copy for scattering amplitudes relates gravity to Yang-Mills amplitudes, it is appreciated by the community that most prescriptions for the classical double copy relate classical gravitational solutions to electromagnetic ones. 
A classical double copy prescription for mapping gravity to non-abelian gauge theory remains elusive. Progress in understanding this discrepancy has been made in \cite{Bahjat-Abbas:2020cyb}, which pointed out how  abelian and non-abelian gauge theory objects may be associated with the same object on the gravity side. Furthermore, by studying gravity as the convolution of left and right Yang-Mills theories \cite{Anastasiou:2014qba}, one may obtain a ``convolutional double copy" which maps gauge and gravity solutions in the manifest Lorenz and de-Donder gauges \cite{Luna:2020adi}. The scenarios studied in the present work considered gravity solutions that, under the Newman-Penrose map, correspond to electromagnetic field configurations. Moving forward, it would be interesting to understand how the Newman-Penrose map may be used to relate gravitational solutions to non-abelian gauge field configurations. The many-to-one nature of the Newman-Penrose map hints that this could be a natural language to address such questions. 

With a few exceptions  \cite{Luna:2018dpt,Luna:2020adi}, all classical double copy prescriptions, including the one presented here, rely on certain coordinate and gauge choices. Yet the relationship between diffeomorphism invariance and gauge redundancy under the double copy remains somewhat obscure. 
In particular, the number of gauge redundancies is different from the number of redundancies associated with diffeomorphism invariance. Once again the novel way in which the Newman-Penrose map projects out ambiguous degrees of freedom, suggests that this might be a promising framework in which to study such questions. Another possible strategy would be to study the behavior of manifestly gauge and diffeomorphism invariant objects under the double copy. On the gauge theory side, Wilson loops, or more generally, spin networks, are natural, gauge invariant quantities with clear gravitational analogues. Recently, progress along this direction was made by using gravitational Wilson line operators, defined in terms of the linearized graviton field, to match certain topological information on both sides of the double copy in a way that is gauge invariant  \cite{Alfonsi:2020lub}. It would be interesting to study such properties using the Newman-Penrose map where it may be natural to construct coordinate invariant gravitational Wilson lines (constructed out of the connection) and study their behavior under the map.

By introducing a novel map between gauge and gravity solutions, this work has opened up new avenues for exploration of the classical  correspondence between the two types of theories.

\section*{Acknowledgements}
We thank Chris White for useful discussions. We also thank  Steve Carlip, Ricardo Monteiro, Donal O'Connell and Chris White for comments on the draft. KF thanks Ond\v{r}ej Hul\'{i}k for helpful discussions and comments.
GH is grateful to Jim Gates for stimulating discussions. GE is supported by the U.S. Department of Energy, under grant number DE-SC0011637.
GE thanks the Berkeley Center for Theoretical Physics and Lawrence Berkeley National Laboratory for their hospitality during the completion of this work. The work of KF was supported by the European Structural and Investment Fund and the Czech Ministry of Education, Youth and Sports (Project CoGraDS - CZ.02.1.01/0.0/0.0/15_003/0000437). The work of MG is supported by the LDRD program at Los Alamos National Laboratory and by the U.S. Department of Energy, Office of High Energy Physics, under Contract No. DE-AC52-06NA25396. 

\newpage
\appendix

\section{Spinors, Kerr's Theorem, and Kerr-Schild Backgrounds}
\label{sec:KerrThm}
Tensors can be more fundamentally described using spinors. We briefly review this formalism here,
closely following
\cite{tod1982self} and \cite{Huggett:1986fs}. We then give an outline of the derivation of Kerr's Theorem for a Kerr-Schild background, which was critical to the set-up of the Newman-Penrose map. A proof of Kerr's theorem using the Newman-Penrose formalism can be found in \cite{1976CMaPh..47...75C} and \cite{doi:10.1063/1.523388}. A proof using spinors can be found in 
\cite{Huggett:1986fs}, which we follow in this appendix.
Here, as in the main text, we restrict to four real spacetime dimensions (for a generalization to higher dimensions see \cite{Monteiro:2018xev}).

\subsection{Spinors}
\label{App:spinors}
With Cartesian coordinates $(t,x,y,z)$ in flat four-dimensional Minkowski space, we form the light-cone coordinates
 $(u,v,\zeta,\bar{\zeta})$ given 
 by \eqref{eq:coor}, with the flat-space metric 
\begin{equation}
    ds^2_0 = 2 (du dv- d\zeta d \bar{\zeta})\,.
\end{equation}

Any vector $V$ can be written in the familiar matrix form as  
$V^\mu= V^{A\dot{A}} \sigma^\mu_{A \dot{A}} $, where $\sigma^\mu=(1,\vec{\sigma})$ and $\vec{\sigma}$ are the Pauli matrices, giving an invertible map between vectors and $2\times2$ matrices, 
\begin{eqnarray}
   V&=&V^\mu \frac{\partial}{\partial X^\mu}=V^{A \dot{A}} 
   \partial_{A \dot{A}}  \,  \quad
   \therefore \,\, V^\mu \longleftrightarrow V^{A\dot{A}} . 
   \label{eq:Vaadot}
\end{eqnarray}
$\SLC$ transformations on the spinors induce the usual $2\!\!:\!\!1$ map onto the group of
proper orthochronous Lorentz transformations acting on the associated vector. 
Spinor indices are raised and lowered using the $\SLC$ invariants $\epsilon_{AB}$ and $\epsilon_{\dot{A} \dot{B}}$, such that 
\begin{eqnarray}
\nonumber
\alpha_A = \alpha^B\epsilon_{BA}\,, \qquad \alpha^A &=&\epsilon^{AB} \alpha_B\,,
\end{eqnarray}
where $\epsilon^{AB} \epsilon_{CB}=\delta^A_C$, and $\epsilon_{01} = 1 = -\epsilon^{01}$.
The $\SLC$ invariant inner product between two spinors $\alpha$ and $\beta$ is given 
by $\epsilon_{AB} \alpha^A \beta^B$. 

Using this map, any vector $V^{A\dot A}$ (null or non-null) can be expanded in terms of a basis of commuting spinors $(o_A,\iota_A)$ called a 
{\em dyad}. The dyad is normalized as 
$o_B \iota^B=o^A \epsilon_{AB}  \iota^B=1$ (which also means that $o^A o_A = 0 = \iota^A \iota_A$), and similarly for spinors with dotted indices. The normalized dyad 
satisfies 
\begin{equation}
    o_A \iota_B- \iota_A o_B = \epsilon_{AB}~.
    \label{appeq:M4metric}
\end{equation}
This form, and the analogous one 
for $\bar{\epsilon}_{{\dot A} {\dot B}}$, is 
preserved by $\SLC$ transformations 
on $(o_A,\iota_B)$ (and for complex spacetimes, by independent $\SLC$ transformations 
on $(o_A,\iota_B)$ and
$(\bar{o}_{\dot A},\bar{\iota}_{\dot B})$). 

Then the matrix component \eqref{eq:Vaadot} may be written as
\begin{equation}
    V^{A\dot A}= c_1  o^A \bar{o}^{\dot A}+c_2 o^A \bar{\iota}^{\dot A}+c_3 \iota^A \bar{o}^{\dot A}+c_4  \iota^A \bar{\iota}^{\dot A}\,,
\end{equation}
where the $c_i$'s can uniquely be obtained from $V^{A\dot A}$ by contracting with any two members of the dyad and using the dyad normalization condition, yielding
\begin{eqnarray}
    X^{A\dot A}
     &=& v o^A \bar{o}^{\dot A}+\zeta o^A \bar{\iota}^{\dot A}+\bar{\zeta} \iota^A \bar{o}^{\dot A}+u \iota^A \bar{\iota}^{\dot A}  \,, \\
     \nabla_{A \dot A}
     &=& o_A \bar{o}_{\dot A}\Delta - o_A \bar{\iota}_{\dot A} 
     \bar{\delta} - \iota_A \bar{o}_{\dot A} \delta +  \iota_A \bar{\iota}_{\dot A} D\, ,
     \label{eq:XandPartialspinors}
\end{eqnarray}
where the directional derivatives $D, \Delta, \delta$ and $\bar{\delta}$ are defined 
in \eqref{directional-derivatives}.
We consider a (fixed) canonical basis (denoted hereafter by a superscript 
`(0)')
\begin{equation} 
o^{(0)A}  = \begin{pmatrix} 1 \\ 0 \end{pmatrix}, \quad
\iota^{(0)A} = \begin{pmatrix} 0 \\ 1 \end{pmatrix}, 
\end{equation}
so that \eqref{eq:XandPartialspinors} may be written as  
\begin{eqnarray}
    X^{A\dot A} =
    \begin{pmatrix}
     v & \zeta \\ 
     \bar{\zeta} & u \end{pmatrix} 
     , \quad
    \partial_{A \dot A}  = 
    \begin{pmatrix}
     \partial_v & \partial_{\zeta} \\ 
     \partial_{\bar{\zeta}} & \partial_u 
    \end{pmatrix} .
\end{eqnarray}
Expanding, we find that
\begin{eqnarray}
    \partial_{A \dot A} =  
    o^{(0)}_A \bar{o}^{(0)}_{\dot A}\partial_u - o^{(0)}_A \bar{\iota}^{(0)}_{\dot A} 
    \partial_{\bar{\zeta}} - \iota^{(0)}_A \bar{o}^{(0)}_{\dot A} \partial_\zeta+ \iota^{(0)}_A \bar{\iota}^{(0)}_{\dot A} \partial_v \,.
\end{eqnarray}
The rank of a null vector is one, so that it can be written in terms of the outer product of two spinors $\alpha$ and $\beta$ if complex, and a single spinor if real. That is, 
\begin{eqnarray}
    k^{A\dot{A}} & =& \alpha^A \bar{\alpha}^{\dot{A}} \quad \text{for} \,\, \hbox{$k$ real},  \\ 
    m^{A\dot{A}} & =& \alpha^A \bar{\beta}^{\dot{A}}  \quad \text{for} \,\, \hbox{$m$ complex}. 
\end{eqnarray}
In terms of the fixed canonical dyad $(o^{(0)}_A, \iota^{(0)}_A)$, a tetrad of null vectors $(l,n,m,\bar{m})$ is then 
given by 
 \begin{eqnarray}
 \nonumber
 l &= l^{A\dot A} \partial_{A \dot{A}} &= \partial_v \,, \quad \bar{m}  = m^{A\dot A} \partial_{A \dot{A}} = \partial_{\bar{\zeta}}\,,\\ 
 n &= n^{A\dot A} \partial_{A \dot{A}} &= \partial_u
 \,,\quad  m = m^{A\dot A} \partial_{A \dot{A}} = \partial_\zeta \,.
 \end{eqnarray}
Here we have used \eqref{eq:XandPartialspinors} and identified   
\bea
 l^{A\dot A} = o^{(0)A} \bar{o}^{{(0)}\dot A}\,, 
 \qquad n^{A\dot A} = \iota^{(0)A} \bar{\iota}^{{(0)}\dot A}\,,
 \qquad m^{A\dot A}=  o^{(0)A} \bar{\iota}^{{(0)}\dot A}\,,
 \qquad \bar{m}^{A\dot A} = \iota^{(0)A} \bar{o}^{{(0)}\dot A} \,,
 \label{app:vector-defn}
\eea
i.e. such that \eqref{eq:tentradnorm} is maintained, and that $l$ is future null pointing.

\subsection{Kerr's Theorem}
\label{app:Kerr-Theorem-KS}
A cogent discussion of Kerr's theorem may be found in \cite{Huggett:1986fs}, which we summarize here. In what follows, we consider 
general spacetime dependent dyads for Minkowski space and solve the geodetic and shear-free conditions 
that directly lead to Kerr's theorem.

Requiring that $l$ be tangent to a null geodetic congruence implies 
\begin{equation}
  o^A D o_A =
  o^A o^B \bar{o}^{\dot{B}} \nabla_{B \dot{B}} o_A 
  =0\,. 
  \label{eq:nullgeodeticspinorform}
\end{equation}
The shear $\sigma$ of $l$ is given by
\begin{equation}
\sigma= m^\mu m^\nu \nabla_\mu l_\nu = o^A o^B \bar{\iota}^{\dot B} \nabla_{B \dot{B}} o_A \,.
\end{equation}
Taken together, requiring that $l$ be geodetic and 
shear-free implies the single equation \cite{Huggett:1986fs}
\begin{equation}
    o^A o^B \nabla_{B \dot{B}} o_A=0\, .
    \label{gsf-eqn}
\end{equation}

Since a re-scaling of $o_A$ by an arbitrary function still gives a null and shear-free $l$, 
we can, without loss of generality fix the normalization of $o_A$, which for a normalized dyad then fixes that of $\iota$. Then starting with the dyad $(o^{(0)}_A, \iota^{(0)}_A)$ describing the 
light-cone coordinate null tetrad of \ref{App:spinors}, the only possible solution for $o_A$ preserving \eqref{appeq:M4metric} is a local null Lorentz transformation for $o_A$ or $\iota_A$. Since the latter does not change the repeated principal null direction $l=\partial_v$, we consider the former.
The most 
general expression for the dyad describing Minkwoski space and an 
associated null and shear-free ray is then given by  
\cite{1979GReGr..10..273K}
\begin{eqnarray}
    o_A & = & o^{(0)}_A - \bar{\Phi}(x) \iota^{(0)}_A, \quad
    \iota_A  =  \iota^{(0)}_A, 
    \label{app:general-dyad}
\end{eqnarray}
where $\bar{\Phi}$ is thus far an undetermined function. The null tetrad \eqref{app:vector-defn}
for this dyad can then be obtained either by
the rules for 
null Lorentz transformations \cite{Chandrasekhar:1985kt} or by direct substitution. 
Either way one finds
\vspace{-0.1 in}
\bea
\nonumber
l &=& \partial_v - \Phi\partial_\zeta - \bar{\Phi}\partial_{\bar{\zeta}} + \Phi\bar{\Phi}\partial_u\, ,  \\ \nonumber
n &=& \partial_u \,, \\ \nonumber
m &=& \partial_{\zeta} - \bar{\Phi}\partial_u \, , \\
\bar{m} &=& \partial_{\bar{\zeta}} - \Phi\partial_u\, ,
\label{App:null-tetrad}
\eea
and associated dual 1-forms 
\begin{eqnarray}
\nonumber
\bm{\omega}^1_0 &= & \iota_A \bar{\iota}_{\dot A} d x^{A \dot A} = dv \,,
\label{eq:flat-space-tetrad-1} \\ \nonumber
\bm{\omega}^2_0 &=& o_A \bar{o}_{\dot A} d x^{A \dot A}=  du+ \bar{\Phi}d\zeta + \Phi d\bar{\zeta} + \Phi\bar{\Phi}dv\, , \label{eq:flat-space-tetrad-2} \\ \nonumber
\bm{\omega}^3_0 &=& -\iota_A \bar{o}_{\dot A} d x^{A \dot A} =  \Phi dv + d\zeta \,, 
\label{eq:flat-space-tetrad-3} \\ 
\bm{\omega}^4_0 &=& -o_A \bar{\iota}_{\dot A} d x^{A \dot A}= \bar{\Phi}dv + d\bar{\zeta}\,,
\label{eq:flat-space-tetrad-4}
\end{eqnarray}
with the metric
\vspace{-0.1 in}
\begin{eqnarray} 
g_0 &=& 2(\bm{\omega}^1_0\bm{\omega}^2_0 - \bm{\omega}^3_0\bm{\omega}^4_0)\,.
\label{app:Minkowski-metric}
\end{eqnarray}

The function $\Phi$ cannot be arbitrary, since $l$ is constrained to be null and shear-free. 
Upon direct substitution of this form for the dyad into the geodetic and shear-free equation 
(\ref{gsf-eqn}), and using the derivative for flat-space, one obtains (\ref{PDEs}), namely \cite{Huggett:1986fs}
\begin{eqnarray}
\nonumber
0 &=& o^A o^B \nabla_{B \dot{B}} o_A  \\ \nonumber
&=& (o^{(0)A}- \bar{\Phi}(x) \iota^{(0)A})( o^{(0)B}- \bar{\Phi}(x) \iota^{(0)B})
\partial_{B \dot{B}} (o^{(0)}_A - \bar{\Phi}(x) \iota^{(0)}_A)  \\ \nonumber
&=& ( o^{(0)B}- \bar{\Phi}(x) \iota^{(0)B})
 \partial_{B \dot{B}} \bar{\Phi}(x)  \\  
 &=& (\partial_\zeta \bar{\Phi} - \bar{\Phi} \partial_u \bar{\Phi} ) \bar{o}_{\dot{B}} 
 + (- \partial_v \bar{\Phi} + \bar{\Phi} \partial_{\bar{\zeta}}\bar{\Phi}) \bar{\iota}_{\dot{B}}
\end{eqnarray}
or equivalently 
\begin{equation}
\Phi_{,v} = \Phi\Phi_{,\zeta}\,, \qquad \Phi_{,\bar{\zeta}} = \Phi\Phi_{,u}\, .
\label{app:nonlinearPDEs}
\end{equation}
The general solution to this equation is given by an arbitrary analytic function $F$, with $\Phi$ defined implicitly 
by the solution to 
\begin{equation}
    F(\Phi, u + \Phi\bar{\zeta}, \zeta + \Phi v)=0 \,.
\end{equation}
A key step in proving this theorem is in first showing that 
\begin{eqnarray}
X_1 &:=& u + \Phi\bar{\zeta}\,, \qquad X_2 :=\zeta + \Phi v  \,,
\end{eqnarray}
are two independent solutions to the two linear partial differential equations 
\begin{eqnarray}
(\partial_v - \Phi \partial_\zeta) X  &=& 0  \,, \qquad
(\partial_{\bar \zeta} - \Phi \partial_u) X =0 \,,
\label{eqn-linearPDES}
\end{eqnarray}
whence $F(\Phi, X_1, X_2)=0$ is a general solution. This integrability condition is
equivalent to requiring that these two linear differential operators 
be closed under commutation, which then implies that $\Phi$ is a solution to 
\eqref{app:nonlinearPDEs} \cite{1976CMaPh..47...75C}.
This relation between arbitrary analytic functions of three arguments and general null shear-free congruences in flat four-dimensional Minkowski space is known as Kerr's Theorem \cite{Penrose:1967wn,1976CMaPh..47...75C} (see also 
\cite{Huggett:1986fs}).

\subsection{From Flat-Space to Kerr-Schild Backgrounds: the Kerr-Wilson Trick}

So far the discussion has been restricted to Minkowski space. 
Surprisingly,  
from 
Kerr's Theorem the SNGC of flat space can be extended to an SNGC of a Kerr-Schild metric.   
To see this, one presupposes a Kerr-Schild metric with a given null vector $l^\mu$. For vacuum and certain Einstein-Maxwell spacetimes,  
\cite{1969JMP....10.1842D} showed that the Kerr-Schild spacetime is algebraically special 
with $l$ as a principal null direction, so by 
the Goldberg-Sachs theorem $l$ is tangent to an SNGC. It then follows that $l$ is also tangent to a null, SNGC with respect to the Minkowski metric. Therefore the assumptions of Kerr's theorem hold, so there exists some analytic function $F$ whose zero set generates $l$.

Next, one may use the following Kerr-Wilson trick to express the Kerr-Schild 
metric in terms of $\Phi$ \cite{1979GReGr..10..273K,doi:10.1063/1.523388}.
It is a simple 
matter of first writing the Minkowksi metric in terms of the 1-forms 
\eqref{eq:flat-space-tetrad-1} generated by the dyad \eqref{app:general-dyad} given by $\Phi$,  yielding  \eqref{app:Minkowski-metric}. 
Then a null tetrad 
for the {\em Kerr-Schild background} can be obtained by shifting
the dual $\bm{\omega}^1_0$, namely 
\begin{eqnarray} 
\nonumber
\bm{\omega}^i &=& \bm{\omega}^i_0 \,, \quad i=2,3,4\,, \\
\bm{\omega}^1 &=& \bm{\omega}^1_0+ \tfrac{1}{2}V\bm{\omega}^2 = dv + \tfrac{1}{2}V\bm{\omega}^2 \,,
\end{eqnarray} 
and a corresponding shift in $n$
\begin{equation} 
n= n_0 - \tfrac{1}{2} Vl\,,
\end{equation} 
while the other vectors are held fixed. This shift ``completes the square" of the Kerr-Schild metric in this basis \cite{1979GReGr..10..273K}. Note that the construction described here, in terms of a null rotation 
on the $(o^{(0)}_A, \iota^{(0)}_A)$  dyad, explains the origin of the initially odd-looking expressions (\ref{tetrad vectors}) for the null tetrad $(l,n,m,\bar{m})$ in terms of $\Phi$. 

For certain Einstein-Maxwell spacetimes, the above argument carries 
through due to a generalization of the Goldberg-Sachs theorem. 
If
the electromagnetic field strength tensor is \emph{null} then the 
generalization states that the spacetime is algebraically special. For more details on generalizations of the 
Goldberg-Sachs theorem, see chapter 7 of \cite{Stephani:2003tm}.
Kerr-Schild spacetimes more general than the classes of solutions to Einstein's equations discussed above can still be related in the same manner to Kerr's theorem, provided that $l$ is tangent to an SNGC.

We conclude this section with an observation. 
The single geodesic and shear-free equation \eqref{gsf-eqn} for $o^A$, expressed in terms of the general dyad \eqref{app:general-dyad} is 
equivalent to the operator
\begin{equation} 
\tilde{\delta}_{\dot{A}}:= o^A\nabla_{A\dot{A}}
\end{equation}
annihilating $\Phi$, namely 
\begin{equation} 
\tilde{\delta}_{\dot{A}}\bar{\Phi} = 0\,.
\label{eq:spinphi}
\end{equation}
As we shall see in the following section, this and related operators play a fundamental role in mapping $\Phi$ to a solution to the source-free self-dual Maxwell equations in flat-space.

\section{Spinorial Realization of the Newman-Penrose Map}
\label{app:khat}
In \cite{tod1982self}, Tod presents an algorithm for associating \emph{null} Maxwell fields in flat-space with \emph{self-dual} Kerr-Schild solutions of general relativity using spinorial methods. Here, we adapt his methods to the Newman-Penrose map, and discuss the spinorial origin of the $\hat{k}$ operator, originally discussed as a method for generating self-dual solutions in \cite{Monteiro:2011pc,Parkes:1992rz}. The notation in this section follows \cite{Newman:2009}.

Let $(l,n,m,\bar{m}) = (\partial_v,\partial_u,\partial_\zeta,\partial_{\bar{\zeta}})$ 
be a null tetrad for the flat metric $g_0 = 2(dudv-d\zeta d\bar{\zeta})$, and let $(o,\iota)$ be a normalized dyad\footnote{In this appendix we will only consider flat-space, so for notational convenience, we suppress the subscript `(0)' on the tetrad and dyad elements.} satisfying $o_A\iota^A = 1$, and 
\be
l^{A\dot{A}} = o^A\bar{o}^{\dot{A}}, \quad n^{A\dot{A}} = \iota^A\bar{\iota}^{\dot{A}}, \quad
m^{A\dot{A}} = o^A\bar{\iota}^{\dot{A}}, \quad
\bar{m}^{A\dot{A}} = \iota^A\bar{o}^{\dot{A}}. \label{squareRootOfTetrad}
\ee
Let us also define the ``spinor directional derivative," 
\be
\delta_{A} = \bar{\iota}^{\dot{A}}\partial_{A\dot{A}} = o_A\partial_u - \iota_A\partial_\zeta\,.
\label{eq:spinordirectionalderivative}
\ee
Given a harmonic complex scalar, $\Phi$, which we may later take to be associated with a Kerr-Schild spacetime \eqref{eq:tetrad} as described in \ref{app:Kerr-Theorem-KS}, one can construct a self-dual Maxwell spinor  given by
\bea
\nonumber
\Phi_{AB} &:=& \delta_{A}\delta_{B}\Phi \\
&=& o_A o_B\Phi_{,uu} - (o_A\iota_B   + \iota_A o_B)\Phi_{,u\zeta} + \iota_A\iota_B\Phi_{,\zeta\zeta}\,.
\label{eq:PhiAB}
\eea

The field strength tensor $F_{A\dot{A}B\dot{B}}$, can be recovered from the Maxwell spinor $\Phi_{\dot{A}\dot{B}}$ via the relation
\be
F_{A\dot{A}B\dot{B}} = \epsilon_{\dot{A}\dot{B}}\Phi_{\stackrel{\,}{A}\stackrel{\,}{B}}\,.
\ee
Applying the definition $\epsilon_{\dot{A}\dot{B}} = \bar{o}_{\dot{A}}\bar{\iota}_{\dot{B}} - \bar{\iota}_{\dot{A}}\bar{o}_{\dot{B}}$, and making use of \eqref{squareRootOfTetrad}, one finds
\bea
F_{A\dot{A}B\dot{B}} &=& (l_{A\dot{A}}m_{B\dot{B}}- m_{A\dot{A}}l_{B\dot{B}})\Phi_{,uu} \nonumber \\ &\,+& (n_{A\dot{A}}l_{B\dot{B}} - l_{A\dot{A}}n_{B\dot{B}} + m_{A\dot{A}}\bar{m}_{B\dot{B}} - \bar{m}_{A\dot{A}}m_{B\dot{B}} )\Phi_{,u\zeta} \nonumber \\ &\,+&   (\bar{m}_{A\dot{A}}n_{B\dot{B}} - n_{A\dot{A}}\bar{m}_{B\dot{B}})\Phi_{,\zeta\zeta}
\eea
which we can rewrite as a two-form $F = \frac{1}{2} F_{A\dot{A}B\dot{B}}dx^{A\dot{A}}\wedge dx^{B\dot{B}}$ which gives
\be
F = -\Phi_{,uu}du\wedge d\bar{\zeta} - \Phi_{,u\zeta}(du\wedge dv + d\zeta \wedge d\bar{\zeta}) + \Phi_{,\zeta\zeta}dv\wedge d\zeta\,, \label{SD field strength 2}
\ee
which agrees with \eqref{SD field strength} up to an overall constant factor. 

Note that we have not yet used the fact that $\Phi$ is harmonic, however we have obtained the same result as we found from the tensorial realization of the Newman-Penrose map. To understand this apparent mismatch, recall that for the tensorial Newman-Penrose map, the field strength two-form is computed as $F = d\hat{k}\Phi$, and so is manifestly closed, but one needs to impose the harmonic condition $\Box_0\Phi = 0$ in order to put it into the form \eqref{SD field strength 2}, from which it follows that $F$ is self-dual. On the other hand, in the spinorial formalism presented here, we obtain \eqref{SD field strength 2} without imposing the harmonic condition, and so the field strength obtained in this way is manifestly self-dual, but is not, in general, closed. Taking the exterior derivative of \eqref{SD field strength 2}, one finds
\be
dF = (\Phi_{,uuv} - \Phi_{,u\zeta\bar{\zeta}})du\wedge dv\wedge d\bar{\zeta} + (\Phi_{,\zeta\zeta\bar{\zeta}} - \Phi_{,uv\zeta})dv\wedge d\zeta \wedge d\bar{\zeta}\,,
\ee
which clearly vanishes when $\Phi$ is harmonic. 
Thus, the spinorial description of the Newman-Penrose map is complementary to the tensorial description, and when the harmonic condition is imposed, we can make use of the Poincar\'{e} lemma to locally integrate $F = dA$ back to a gauge field $A \propto \hat{k}\Phi$, where $\hat{k}$ is given by \eqref{diffOp}  in tensorial form, and can be written in spinorial form as
\begin{align}
\hat{k} =  \frac{Q}{2\pi \epsilon_0}dx^{A\dot{A}}\bar{\iota}_{\dot{A}} \delta_A\,.
\end{align}

While the tensorial description has the advantage of being more familiar to many readers, we see that the spinorial description has its own advantages--- the operator $\hat{k}$ appears somewhat mysterious in the tensorial formulation, however it can be understood in the spinorial approach as an artifact of the ``spinor directional derivative" $\delta_{A}$, which has a relatively natural interpretation. 
And the set of non-linear partial differential equations \eqref{app:nonlinearPDEs} of which $\Phi$ is 
a solution to, has a relatively simple expression in terms of a  spinorial directional derivative, namely \eqref{eq:spinphi}. Note that the choice of spinorial directional derivative \eqref{eq:spinordirectionalderivative} is not unique. Indeed, one may consider a spinorial directional derivative along a general direction in spin space i.e.
\bea
&&\delta_A = \left(c_1 \bar{o}^{\dot{A}} + c_2 \bar{\iota}^{\dot{A}} \right) \partial_{A\dot{A}} \,,
\eea
(and its conjugate $\bar{\delta}_{\dot{A}}$), which may a priori act on $\Phi$ or $\bar{\Phi}$. 
However, restricting to a direction in spin space that leads to an interesting map, namely a non-trivial gauge field that satisfies the self-dual condition \eqref{anti self-dual}, leads to the choice \eqref{eq:spinordirectionalderivative} (and analogously the choice of differential operator in \eqref{diffOp})
when acting on $\Phi$ and it's conjugate when acting on $\bar{\Phi}$. While it would be interesting to discover a deeper understanding of how the particular direction in spin space \eqref{eq:spinordirectionalderivative} is related to the Newman-Penrose map, it is beyond the scope of the current work.

\section{Details of the Kerr Metric} 
\label{app:ConfirmMetric}
In this appendix we present some details of the Kerr metric that are relevant to \ref{subsec:KerrBH}. First we verify that the metric for Kerr (and in the $a\rightarrow 0$ limit, Schwarzschild) can be obtained from the complex scalar $\Phi$ of \eqref{eq:PhiKerr}. This demonstrates the potential for the Newman-Penrose map to be inverted: starting with a self-dual gauge field generated by the operator $\hat{k}$ acting on a complex scalar $\Phi$, we may obtain solutions to the vacuum Einstein equations with only the mass parameter as external input. More general non-vacuum solutions may also be generated, and we hope to explore this in detail in future work. We conclude by deriving the relation of the Kerr metric in Boyer-Lindquist coordinates to the two Kerr-Schild forms \eqref{eq:KerrKS} and \eqref{eqn:Kerr-from-Phi}.

\subsection{Generating the Kerr-Schild Form of the Metric from $\Phi$}
\label{app:ConfirmMetricSubSec}
After making use of the definition of $r$ \eqref{rdef} and $z = r\cos\theta$, $\Phi$ for the Kerr metric \eqref{eq:PhiKerr} can be written
\bea
\Phi  &=& \frac{1}{2\bar{\zeta}}\left((v - u + \sqrt{2}ia)-\sqrt{2}(r+ia\cos\theta)\right) =\frac{r-ia}{\sqrt{2}\bar{\zeta}}\left(\cos\theta -1\right).
\label{eq:PhiApp}
\eea
From this we can define the null one-form
\begin{align}
\bm{\omega}^2 = du + \bar{\Phi} d\zeta + \Phi d\bar{\zeta} + \Phi \bar{\Phi} dv\,.
\label{eq:onefApp}
\end{align}
as well as the function $V$ so that the full metric can then be written in Kerr-Schild form as
\begin{align}
g = g_0 + V \bm{\omega}^2 \otimes \bm{\omega}^2.
\label{eq:metApp}
\end{align}
where $g_0$ is the flat-space metric. In the case of vacuum spacetimes, the real scalar $V$ is also determined by $\Phi$:
\begin{align}
V &= \frac{M(\rho+\bar{\rho})}{P^3}\,,
\end{align}
with 
\begin{align}
\rho = \frac{1+\Phi\bar{\Phi}}{2\bar{\zeta}\Phi - (v-u+\sqrt{2}i a)}\quad{}\textnormal{and}\quad{}
P = \frac{1}{\sqrt{2}} (1+\Phi \bar{\Phi}).
\end{align}
Plugging in \eqref{eq:PhiApp} for the Kerr metric into these equations gives 
\begin{align}
\rho =  \frac{-\sqrt{2}}{(1+\cos\theta)(r+ia\cos\theta)}\quad{}\textnormal{and}\quad{}P= \frac{\sqrt{2}}{1+\cos\theta} \,,
\end{align}
where we made use of the fact that $2\zeta \bar{\zeta} = (r^2+a^2)(1-\cos^2\theta)$. This means the real function $V$ for the Kerr metric is
\begin{align}
V &=- \frac{M r (1+\cos\theta)^2}{(r^2+a^2\cos^2\theta)}\,.
\end{align}
For the null one-form $\bm{\omega}^2$, we can plug \eqref{eq:PhiApp} into \eqref{eq:onefApp} to get
\begin{align}
\bm{\omega}^2 = du + \frac{r-ia}{2\bar{\zeta}} \left( \cos\theta - 1\right)d\bar{\zeta} + \frac{r+ia}{2\zeta } \left( \cos\theta -1\right)d\zeta + \frac{1-\cos\theta}{1+\cos\theta}dv\,.
\end{align}
Transforming to Cartesian coordinates via \eqref{eq:coor}, this becomes:
\begin{align}
\bm{\omega}^2
&=  \frac{\sqrt{2}}{1+\cos\theta}\bigg[dt -\frac{z}{r} dz- \frac{rx+ay}{(r^2+a^2)}dx-\frac{ry-ax}{(r^2+a^2)}dy\bigg]
\end{align}
such that \eqref{eq:onefApp} becomes
\begin{align} 
ds^2&=ds^2_0 - \frac{2 M r}{(r^2+a^2\cos^2\theta)}\left(dt -\frac{z}{r} dz- \frac{rx+ay}{(r^2+a^2)}dx-\frac{ry-ax}{(r^2+a^2)}dy\right)^2 ~.
\label{eq:KerrKS-outgoing-appA}
\end{align}

Note that this form for the Kerr metric 
na\"ively disagrees with that given
in \eqref{eq:KerrKS}, which for ease of direct comparison we reproduce here: 
\begin{align} 
ds^2&=ds^2_0 - \frac{2 M r}{(r^2+a^2\cos^2\theta)}\left(dt +\frac{z}{r} dz+ \frac{rx+ay}{(r^2+a^2)}dx+\frac{ry-ax}{(r^2+a^2)}dy\right)^2.
\label{eq:KerrKS-appA}
\end{align}
However, these two metrics are equivalent, at least in the region 
exterior to the outer horizon $r_+$ \cite{Boyer:1966qh}. For the two pairs of $x$ and $y$ coordinates 
appearing in each of 
these expressions are not the same quantities, as they are each related to more standard coordinates for the Kerr metric 
by a different
coordinate change.

\subsection{Other Coordinate Systems}
\label{app:Kerr-metric-many-forms}

A standard textbook expression for the Kerr metric, in Boyer-Lindquist coordinates (see for instance Chandrasekhar \cite{Chandrasekhar:1985kt}), is given by
\begin{eqnarray} 
ds^2 &=& \frac{\Delta}{\rho^2}(dT-a \sin ^2 \theta d \varphi)^2 - \frac{\sin ^2 \theta}{\rho^2}\left(a dT - (r^2+a^2) d \varphi\right)^2 
- \frac{\rho^2}{\Delta} dr^2 - \rho^2 d \theta^2 ~. 
\label{Kerr-metric:Boyer-Lindquist}
\end{eqnarray}
where 
\begin{eqnarray}
\rho^2 &=& r^2 + a ^2 \cos^2 \theta\,, \qquad \Delta = r^2 - 2Mr +a^2.
\end{eqnarray}
The metric is invariant under a combination of time reversal and a flip of the direction of rotation, but not either transformation individually.
In \eqref{Kerr-metric:Boyer-Lindquist}, we can freely choose either sign 
for $a$ by doing a coordinate transformation $T \rightarrow - T$ or  
$\varphi \rightarrow - \varphi$.  Assume 
we've done that, then we proceed with a fixed sign for $a$ in what follows, except where 
noted at the end of this discussion.
Following \cite{Chandrasekhar:1985kt},  
we consider the ingoing and `$(+)$' and outgoing `$(-)$' null rays\footnote{These are usually defined either as $dv$ or $du$ for the $+$ or $-$ sign respectively, but we use the letter $w$ here for full generality.}
\begin{equation}
dw = dT \pm \frac{r^2 +a ^2}{\Delta} dr ~, \qquad
d \phi = \pm d \varphi + \frac{a}{\Delta} dr ~.
\end{equation}
Only the lower choice `$(-)$' was considered in \cite{Chandrasekhar:1985kt}, but as we shall see, the other 
choice is important as well. These two choices 
will lead to the two different expressions of the Kerr-Schild form for the Kerr metric.
Looking ahead to \eqref{Kerr-Metric:intermediate-KS-form} giving an 
intermediate form of the metric in this new coordinate system, one 
indeed sees that on $d w=d \phi=d \theta=0$ the metric is null. 
Continuing, 
then in the new coordinate system
\begin{equation} 
ds^2 = \frac{\Delta}{\rho^2} \left(dw \mp a \sin^2 \theta d \phi\right)^2 
\mp 2 dr dw +2 a \sin ^2 \theta d \phi dr
- \frac{\sin ^2 \theta}{\rho^2} \left(\pm (r^2+a^2) d \phi - a dw\right)^2 
-\rho^2 d \theta^2  ~.
\end{equation}
 Since the only dependence on the metric on $M$ is in $\Delta$ which is linear, in these coordinates 
the dependence of the metric on $M$ is linear, so the metric is now in the Kerr-Schild form. All the other terms have to give the Minkowski metric. Substituting 
for $\rho$ and $\Delta$ one has 
\begin{eqnarray}
ds^2 &=& (dw \mp dr)^2 - dr^2 -\rho^2 d \theta ^2 - (r^2 +a^2) \sin^2 \theta 
d \phi ^2 + 2 a \sin^2 \theta  d \phi dr \nonumber \\
& & - \frac{2 Mr}{r^2 +a^2 \cos^2 \theta}(dw \mp a \sin^2 \theta d \phi)^2 ~.
\label{Kerr-Metric:intermediate-KS-form}
\end{eqnarray}

The first line is Minkowski space in ``oblate spheroid" coordinates. The 
second line is the Kerr-Schild vector part. 
To see that, define Cartesian coordinates \eqref{eq:sph} 
\begin{eqnarray}
x+i y &=& (r+i a )e^{i \phi} \sin \theta ~,~z= r \cos \theta ~.
\end{eqnarray}
Then 
\begin{equation} 
x^2+y^2 =(r^2 + a^2) \sin^2 \theta\,, \quad{}
\frac{x^2+y^2}{r^2+a^2} + \frac{z^2}{r^2} =1 ~,
\end{equation}
and 
\begin{equation} 
dx^2 + dy^2 +dz^2 =dr^2 +\rho^2 d \theta^2 + (r^2 +a^2) \sin ^2 \theta 
d \phi^2 - 2 a \sin^2 \theta dr d \phi~.
\end{equation}
Finally we let 
\begin{eqnarray} 
dt &=& dw \mp dr ~,
\end{eqnarray} 
then 
\begin{eqnarray}
ds^2 &=& ds^2_0 - \frac{2 Mr}{r^2 +a^2 \cos^2 \theta}(dt \pm dr \mp a \sin^2 \theta d \phi)^2 ~,
\end{eqnarray}
which is in the Kerr-Schild form for either $\pm$.
In summary, starting with a standard textbook form for the Kerr metric 
\eqref{Kerr-metric:Boyer-Lindquist}, then 
by transforming to coordinates associated to either of the special principal null directions, the Kerr metric 
can be written in the Kerr-Schild form in two different ways.

Now let's proceed to write the 1-form in the above metric in terms of the $x,y,z$ coordinates. 
Using 
\begin{eqnarray}
xdy-ydx &=& -a \sin^2 \theta dr + (r^2+a^2) \sin^2 \theta d \phi ~,
\end{eqnarray}
one finds after some algebra 
the outgoing ray (upper sign choice) and ingoing ray (lower sign choice)
\begin{eqnarray}
dt \pm dr \mp a \sin^2 \theta d \phi
&=& dt \pm \left( \frac{z}{r} dz + \frac{rx+ay}{r^2+a^2} dx + \frac{ry-ax}{r^2+a^2} dy
\right)~.
\label{Kerr-Schild-one-form}
\end{eqnarray}
We now see that the lower/`$-$' choice for the outgoing ray is the same as the above  
expression \eqref{eq:KerrKS-outgoing-appA} for 
the Kerr-Schild metric obtained using $\Phi$, and is equivalent to 
the expression in \cite{Chandrasekhar:1985kt} after using slightly different 
coordinate transformations.\footnote{To see that, 
first note that for $d \phi$ in \cite{Chandrasekhar:1985kt}
has an opposite sign choice to the one used here, so 
define 
$d \phi = d \varphi - \frac{a}{\Delta} dr$ and also $x+iy=(r-ia)e^{-i \phi}\sin\theta$, whose only effect is to change the 
sign of $a$ in the above expression.}  The upper/`$+$' choice for the 
incoming ray gives the usual Kerr-Schild 
form of Kerr metric found in the double copy literature and given 
above in \eqref{eq:KerrKS-appA}.

\newpage

\bibliographystyle{JHEP}
\bibliography{NPDC}

\end{document}